\documentclass[preprint,5p,authoryear,breaklinks]{elsarticle}

\usepackage[latin1]{inputenc}
\usepackage[english]{babel}
\usepackage{fontenc}
\usepackage{graphicx}

\usepackage{natbib}
\usepackage{caption}
\DeclareCaptionLabelFormat{fig_lblformat}{\textbf{Fig. #2.}}
\DeclareCaptionLabelFormat{tbl_lblformat}{\textbf{Table #2}\\}
\usepackage{amsmath}
\usepackage{amsfonts}
\usepackage{amssymb}

\usepackage{multicol}

\begin{document}

\title{Long-term \& large-scale viscous evolution of dense planetary rings.}

\author[cea]{J.~Salmon\corref{cor1}}
\ead{julien.salmon@cea.fr}
\author[cea]{S.~Charnoz}
\ead{charnoz@cea.fr}
\author[damtp]{A.~Crida}
\ead{crida@oca.eu}
\author[cea]{A.~Brahic}
\ead{brahic@cea.fr}

\cortext[cor1]{Corresponding author}

\address[cea]{UMR AIM, Universit\'e Paris Diderot/CEA/CNRS, CEA/SAp\\Orme des
Merisiers, bat. 709, 91191, Gif-Sur-Yvette Cedex, FRANCE}
\address[damtp]{Department of Applied Mathematics and Theoretical Physics\\University of Cambridge,
Centre for Mathematical Sciences, Wilberforce Road, Cambridge CB3 0WA, UK}

\begin{abstract}
Planetary rings are common in the outer Solar System but their origin and
long-term evolution is still a matter of debate. It is well known that viscous
spreading is a major evolutionary process for rings, as it globally
redistributes the disk's mass and angular momentum, and can lead to the disk's
loosing mass by infall onto the planet or through the Roche limit. However,
describing this process is highly dependent on the model used for the viscosity.
In this paper we investigate the global and long-term viscous evolution of a
circumplanetary disk. We have developed a simple 1D numerical code, but we use a
physically realistic viscosity model derived from N-body simulations \citep{dais01}, and
dependent on the disk's local properties (surface mass density, particle size,
distance to the planet). Particularly, we include the effects of gravitational
instabilities (wakes) that importantly enhance the disk's viscosity. This method
allows to study the global evolution of the disk over the age of the Solar
System.

Common estimates of the disk's spreading time-scales with constant viscosity
significantly underestimate the rings' lifetime. We show that, with a realistic
viscosity model, an initially narrow ring undergoes two successive evolutionary
stages: (1) a transient rapid spreading when the disk is
self-gra\-vi\-ta\-ting, with
the formation of a density peak inward and an outer region marginally
gravitationally stable, and with an emptying time-scale proportional to
$1/M_0^2$ (where $M_0$ is the disk's initial mass) (2) an asymptotic regime
where the spreading rate continuously slows down as larger parts of the disk
become not-self-gra\-vi\-ta\-ting due to the decrease of the surface density,
until
the disk becomes completely not-self-gra\-vi\-ta\-ting. At this point its
evolution
dramatically slows down, with an emptying time-scale proportional to $1/M_0$,
which significantly increases the disk's lifetime compared to the case with
constant viscosity. We show also that the disk's width scales like $t^{1/4}$
with the realistic viscosity model, while it scales like $t^{1/2}$ in the case
of constant viscosity, resulting in much larger evolutionary time-scales in our
model. We find however that the present shape of Saturn's rings looks like a 100
million-years old disk in our simulations. Concerning Jupiter's, Uranus' and
Neptune's rings that are faint today, it is not likely that they were much more
massive in the past and lost most of their mass due to viscous spreading alone.
\end{abstract}


\maketitle

\section{Introduction}
Saturn's rings are one of the most puzzling object of the Solar System. They've
been studied for over 400 years since their discovery by Galileo in the early
17th century, but still many key questions remain unanswered. In particular, how
old the rings are is still a matter of debate. While published scenarios for
their origin suggest an early formation in the history of the Solar System: (1)
remnant of Saturn's sub-nebula disk \citep{poll73}, (2) tidal disruption of a
comet \citep{done91,char09a}, (3) destruction of a satellite inside Saturn's
Roche zone \citep{poll73, harr84,char09a}, observations and theoretical
arguments such as the viscous spreading of the A ring in a few hundred million
years \citep{espo86}, or the low meteoritic pollution of the rings
\citep{cuzz98}, lead to the conclusion that they must be quite young. Thus we
are in a paradoxical situation.

The rings' evolution is regulated through 3 main physical processes. (1) Viscous
spreading: particle collisions dissipate energy while conserving the total
angular momentum, resulting in the spreading of the disk with the mass being
transferred inward and the angular momentum outward \citep{lynd74, gold82}, (2)
Interactions with the planet's satellites at resonances \citep{gold79}, and (3)
Meteoritic bombardment \citep{cuzz98}. All these processes modify the disk's
angular momentum and transport material, which can thus leave the disk when it
reaches the planet's radius or crosses the Roche limit where it can start to be
accreted into small moons \citep[][\textit{In press}]{char10}. These processes
can then cause a drop of the disk's mass. Seeing how they strongly influence the
possible lifetime of the rings, the paradox stated above must come from an
insufficient understanding of these processes on the long term and from a lack
of models. A more detailed and global study of these processes over long
time-scales is needed.

While process (3) is specific to Saturn's rings, processes (1) and (2) are
common to other disks in the universe, for instance protoplanetary disks,
accretion disks, or galaxies. But Saturn's rings are particular in the sense
that they are devoid of gas and composed only of macroscopic particles, so that
pressure and radiative effects can be neglected compared to gravitational
effects. Consequently, while the approach remains valid, results on the
evolution of a gaseous disk under the aforementioned processes cannot be
directly transposed to Saturn's rings but have to be studied considering their
specificities.

The viscous spreading of a disk has already been studied, particularly accretion
disks \citep{lynd74, bath81, prin81}. However, this process is highly dependent
on the model used to describe the disk's viscosity. Several important works have
been published on local properties of the rings \citep{wisd88, salo92, rich94,
dais99, ohts00, dais01}, resulting in precise estimates of the viscosity
dependence on the disk's parameters (surface density, particle properties, ...),
but these local properties have not been studied yet on large spatial and
temporal scales. In this paper we focus our study on the viscous spreading of
the disk considering a physically realistic viscosity model and the evolution of
the whole disk. Eventually, constrain the evolution of the rings will require to
consider many different physical processes, but this is for now well beyond
current computer capacity. Moreover, questions such as the spreading rate of the
rings, the amount of material falling onto the planet or leaving the rings
through their outer limit, are important issues that still need to be
addressed. 

Published estimates of the rings' viscous age are viscous time-scales $\Delta
r^2/\nu$ that give the time needed for a ring with constant viscosity $\nu$ to
reach the width $\Delta r$. However, we can expect the viscosity of the rings to
be a rapidly varying function of the disk's local properties (surface mass
density, ring particle's radius, distance to the planet), so that such
time-scales might be substantially inaccurate.

In this work we study the effects of a time-dependent and space-dependent
viscosity model on the long term viscous evolution of the complete ring system.
We perform 1-dimensional numerical simulations, along the radial direction, of
the disk's viscous evolution. This implies simplifications of the physics but
allows for simulations over many dynamical times (typically the age of the Solar
System). As a consequence, our results will be a first-order study of the global
viscous evolution of dense planetary rings. However, it may give us clues to
understand what has led to the rings we can observe today. Moreover, this
approach has already been used with important results in the study of
protoplanetary disks \citep[e.g.][]{take96, alib05}.

In section \textbf{2} we describe the viscosity model and the numerical
procedure. In section \textbf{3} we study the viscous evolution  of an initially
narrow ring over 5 billion years using two models: constant viscosity
(hereafter CV) and non constant viscosity (hereafter NCV), and develop
analytical time-scales to describe the disk's evolution. In section \textbf{4}
we analyse the influence of ring parameters. In section \textbf{5} we summarize
the results and discuss the evolution of today's rings of Saturn in the light of
these results.

\section{Numerical method and viscosity model}
In this section we present the hydrodynamical equations describing the evolution
of the rings, the viscosity model we use, and the principles of the numerical
code we have developed. While adopting a very general approach, we present our
results for the case of Saturn's rings.

\subsection{Basic equations}
We use a 1-dimensional approach to study the viscous evolution of a Keplerian
pressureless disk. To compute the evolution of the surface density $ \Sigma
\left( R, t \right) $ of an elementary annulus of the rings located at distance
R from the planet and at time t, we use the same approach as \cite{prin81}. The
equation of mass conservation reads:
\begin{equation}
R \frac{\partial \Sigma}{\partial t} + \frac{\partial}{\partial R} \left( R
\Sigma v_{\rm R} \right) = 0,
\label{masscons}
\end{equation}
where $v_R \left( R, t \right) $ is the radial velocity. The angular
momentum conservation reads:
\begin{align}
\frac{1}{R} \frac{\partial}{\partial R} \left( \nu \Sigma R^3 \frac{\partial
\Omega}{\partial R} \right) = &~\frac{\partial}{\partial t} \left( \Sigma R^2
\Omega \right)\notag\\
 &+ \frac{1}{R}\frac{\partial}{\partial R} \left( \Sigma R^3 \Omega v_{\rm R}
\right),
\label{angmomcons}
\end{align}
where $\Omega = \sqrt{G M/R^3} $ is the orbital frequency, with G the
gravitational constant and M the planet's mass, and $\nu$ is the kinematic
viscosity.
Combining Eqs. (\ref{masscons}) and (\ref{angmomcons}) and replacing $\Omega$ by
its expression we get a single equation for the temporal evolution of the
surface density:
\begin{equation}
\frac{\partial \Sigma}{\partial t} = \frac{3}{R} \frac{\partial}{\partial R}
\left[ \sqrt{R} \frac{\partial}{\partial R} \left( \nu \Sigma \sqrt{R} \right)
\right].
\label{main_equ}
\end{equation}

\subsection{Viscosity model}
\subsubsection{Description of the model}

The viscosity of a disk is related to angular momentum transport through
particle interactions. The angular momentum flux reads \citep{lynd74}
\begin{equation}
 \Phi = 3\pi\Sigma\Omega R^2 \nu,
 \label{ang_mom_flux}
\end{equation}
Using the Boltzmann equation, one can derive this angular momentum flux as
\citep{take01} 
\begin{equation}
\frac{\partial}{\partial t} \left( 2 \pi R \Sigma R U_{\rm \theta} \right) = -
\frac{\partial \Phi}{\partial R}
\end{equation}
where $U_{\rm \theta}$ is the mean tangential velocity averaged azimuthally. The
viscosity can then be computed by estimating the angular momentum flux and
inverting Eq. (\ref{ang_mom_flux}).

Angular momentum transport can be divided in three components, each one related
to a specific viscosity. The translational viscosity $\nu_{\rm trans}$ is
related to the transport of angular momentum due to the random motion of
particles (usually referred to as the ``local'' component, see \cite{gold78}).
Due to their finite size, angular momentum is also transported via sound waves
travelling between the centres of colliding particles (usually referred to as
the ``non-local'' component). This is represented by the collisional viscosity
$\nu_{\rm coll}$ \citep{arak86, wisd88}. Finally, angular momentum is
transported by gravitational scattering of particles due to the presence of
self-gravity wakes \citep{salo92, salo95, rich94, dais99, ohts00}. This is
represented by the gravitational viscosity $\nu_{\rm grav}$.

The self-gravity wakes are gravitational aggregates induced by the effects of
both self-gravity and collisional damping of particles. Outside of a wake,
particles move randomly, but inside their motion becomes coherent, which yields
systematic motion with large bulk viscosity \citep{salo95, dais99}. The wakes
modify the angular momentum transport, and thus the viscosity, through the
gravitational torque \citep{lars84} and the wake motion \citep{lin87}. Two
regimes must then be considered, whether the disk is gravitationally stable (no
wakes) or not. This is measured by the Toomre Q parameter \citep{toom64}
\begin{equation}
 Q = \frac{\Omega \sigma_{\rm r}}{3.36 G\Sigma},
\label{toomreQ}
\end{equation}
where $\sigma_{\rm r}$ is the particle radial velocity dispersion. Even though
Toomre showed that the disk is gravitationally unstable for $Q \lesssim 1$,
N-body simulations by \cite{salo95} showed that wakes start to form for $Q
\lesssim 2$.

In our simulations, we use the following formulations for the viscosity
components. We set the transition between the two regimes,
self-gra\-vi\-ta\-ting and not-self-gra\-vi\-ta\-ting (hereafter SG and NSG),
strictly at $Q =2$.
\begin{equation}
\nu = \nu_{\rm trans} + \nu_{\rm coll} + \nu_{\rm grav},
\end{equation}
with
\begin{equation}
\nu_{\rm trans} = \begin{cases}
\frac{\sigma_{\rm r}^2}{2 \Omega} \left( \frac{0.46 \tau}{1 + \tau^2} \right) &
\mbox{if } Q > 2, \\
\frac{1}{2} 26 {r_{\rm h}^*}^5 \frac{G^2\Sigma^2}{\Omega^3} & \mbox{if } Q < 2,
\\
\end{cases}
\label{nu_trans_eq}
\end{equation}
\begin{equation}
\nu_{\rm coll} = r_{\rm p}^2 \Omega \tau ~~~\forall ~Q, \\
\label{nu_coll_eq}
\end{equation}
\begin{equation}
\nu_{\rm grav} = \begin{cases}
0 & \mbox{if } Q > 2, \\
\nu_{\rm trans} & \mbox{if } Q < 2, \\
\end{cases} \\
\label{nu_grav_eq}
\end{equation}
where $\tau = 3 \Sigma/\left( 4 r_{\rm p} \rho_{\rm p} \right)$ is the optical
depth with $r_{\rm p}$ and $\rho_{\rm p}$ the particle's radius and density.
$r_{\rm h}^*$ is a dimensionless parameter equal to the ratio of the particle's
Hill radius $r_{\rm h}$ to its physical diameter: $r_{\rm h}^*=r_{\rm h}/(2
r_{\rm p})$ with $r_{\rm h}=\left( 2 m_{\rm p}/(3 M_s) \right)^{1/3}R$, and
$m_{\rm p}$ is the particle's mass. It's basically a dimensionless measure of
the distance to the planet, while the optical depth serves as a dimensionless
measure of the disk's surface mass density. 

In the NSG regime ($Q > 2$), the translational viscosity is an analytical result
from \cite{gold78} (Eq. (\ref{nu_trans_eq}), top), and the collisional viscosity
is an analytical result from \cite{arak86} (Eq. (\ref{nu_coll_eq})). In the SG
regime ($Q < 2$) (Eqs. \ref{nu_trans_eq}, bottom \& \ref{nu_grav_eq}, bottom),
we use the results of \cite{dais01}, as they are the first to include the
effects of the wakes in the calculation of the viscosity. They performed
multiple 3D shearing box N-body simulations, including the effects of
self-gravity wakes, with 1m-radius particles. Their set of parameters includes
high and low densities, which is suitable to track the rings' evolution in a
large variety of physical conditions. Note that although the collisional
viscosity is also enhanced in the presence of wakes, its value is about one
order of magnitude smaller than the translational and gravitational viscosity in
that case, according to the simulation results of \cite{dais01}. Thus we use the
same prescription for the collisional viscosity whether the disk is
gravitationally stable or not.

The parameter $r_{\rm h}^*$ is also expressed in \cite{dais01} as $\left|
F_{\rm grav} \right|/\left| F_{\rm coll} \right| = 4 {r_{\rm h}^*}^2$, where
$F_{\rm grav}$ is the self-gravity force between particles, and $F_{\rm coll}$
is the impulsive force exerted on particles during collisions. This can be used
to evaluate the velocity dispersion $\sigma_{\rm r}$. If $r_{\rm h}^* < 0.5$,
$F_{\rm coll}$ dominates and the velocity dispersion is the relative Keplerian
velocity between particles $\sigma_{\rm r} = 2 r_{\rm p} \Omega$. Conversely, if
$r_{\rm h}^* > 0.5$, $F_{\rm grav}$ dominates and the velocity dispersion is
regulated to be the particle's surface escape velocity $\sigma_{\rm r} = \sqrt{G
m_{\rm p}/r_{\rm p}}$ \citep{salo95, dais99, ohts99}. For Saturn's rings we have
$r_{\rm h}^* \gtrsim 0.5$ so in the following we always consider that
$\sigma_{\rm r} = \sqrt{G m_{\rm p}/r_{\rm p}}$. Note that we do not consider
here the effects of thermal conduction that could modify the velocity
dispersion.

Toomre's Q parameter depends on the disk's local properties: it decreases with
surface mass density $\Sigma$ and distance to the planet R. Some parts of
the disk may then be gravitationally stable while others are not, and we can
expect theses regions to undergo distinct evolutions. As the disk evolves, the
surface density will decrease because of the disk's spreading, and the Toomre Q
parameter will increase. Then regions initially self-gra\-vi\-ta\-ting may
become not-self-gra\-vi\-ta\-ting as the disk evolves, leading to local
modifications of the viscosity. Note that with the adopted prescription of the
velocity dispersion, Q is also proportional to $r_{\rm p}$. So a disk with
bigger particles will be more likely to become not-self-gra\-vi\-ta\-ting.

\subsubsection{Viscosity dependence on ring parameters}
\begin{figure}[!h]
\begin{center}
\includegraphics[width=8.5cm]{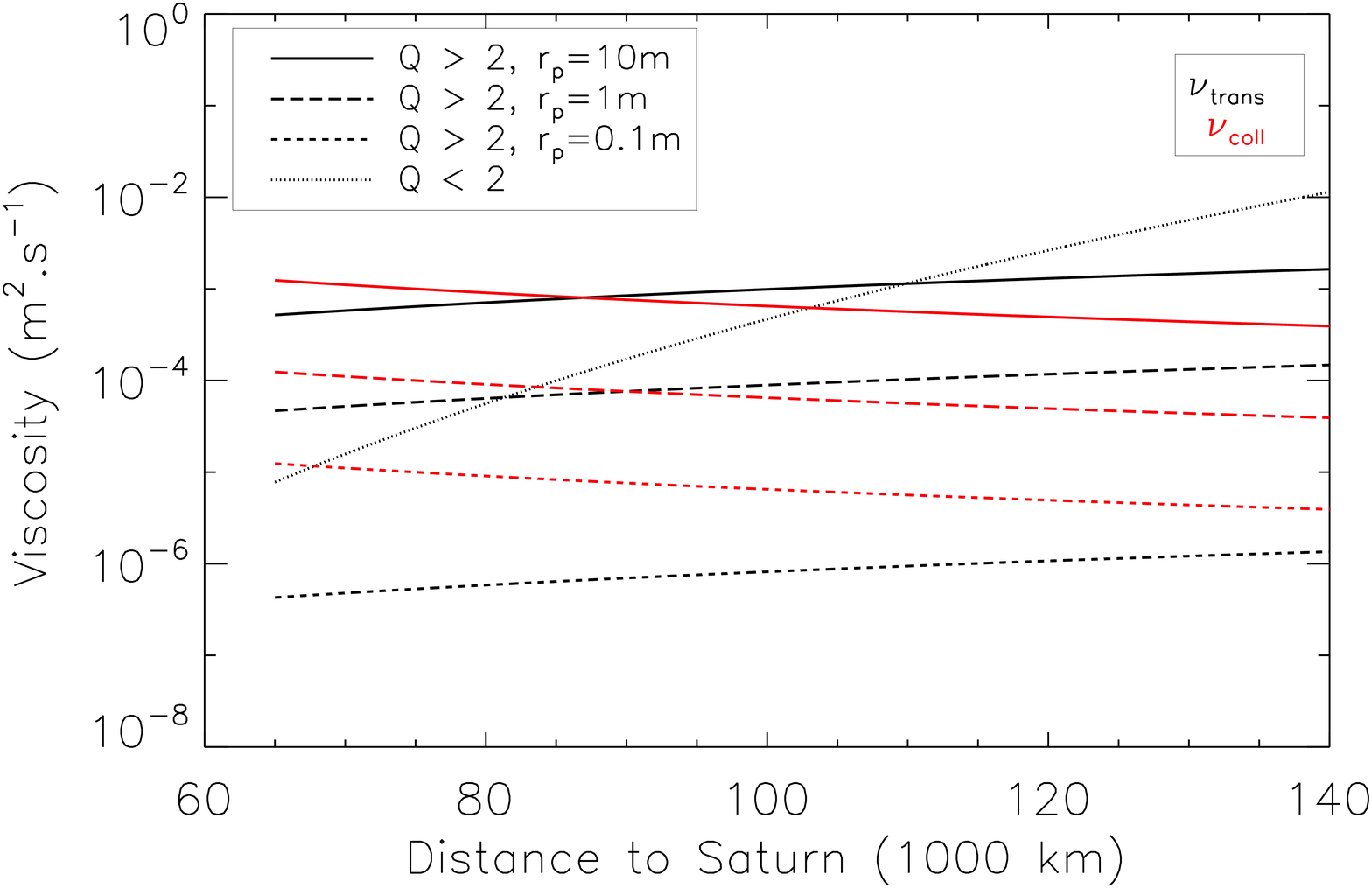}
\end{center}
\caption{Radial evolution of the viscosity for different ring particle
sizes. The surface density is set to $400\rm~kg.m^{-2}$. The translational
viscosity (black curves) increases with the distance to the planet, while the
collisional viscosity (red curves) decreases with the distance. The
translational viscosity can be several orders of magnitude larger in the
self-gra\-vi\-ta\-ting case (black dotted line).}
\label{visco_trans_coll}
\end{figure}

The viscosity components depend on several parameters that are likely to vary
within the rings: particle radius, surface mass density and distance to the
planet (Eqs. (\ref{nu_trans_eq}) and (\ref{nu_coll_eq})). Even though this model
may be used for any planetary ring, we are most interested in Saturn's rings.
Here we discuss the resulting viscosity for the specific case of Saturn. As an
example, we plot in Fig. \ref{visco_trans_coll} the variation of $\nu_{\rm
trans}$ and $\nu_{\rm coll}$, for a constant value of the surface density
\citep[$ 400~\rm kg.m^{-2}$, the surface density of today's A ring, ][]{tisc07},
with respect to R and for different values of $r_{\rm p}$.

Both translational and collisional viscosities increase with particle radius,
but they have opposite dependences on the distance to the planet: the
collisional (translational) viscosity decreases (increases) with R. In the SG
regime, the particle radius does not modify the value of the viscosity itself,
but it impacts Toomre's Q parameter which controls the areas of the rings that
are in the SG regime.

\begin{figure}[!h]
\begin{center}
\includegraphics[width=8.5cm]{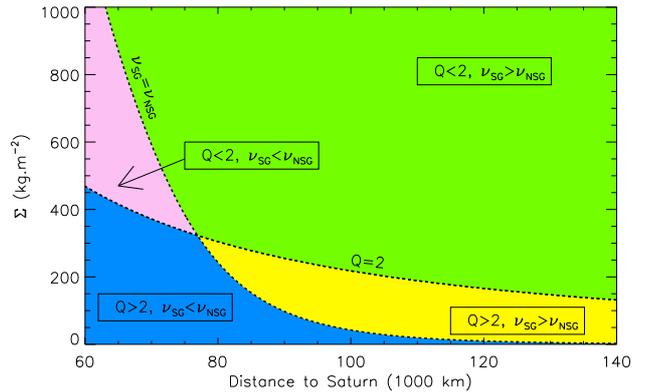}
\end{center}
\caption{Relative magnitude of the total viscosity $\nu_{\rm trans} +
\nu_{\rm coll} + \nu_{\rm grav}$ in the self-gra\-vi\-ta\-ting and
not-self-gra\-vi\-ta\-ting regimes, and position of the regime transition. The
dashed curves are the transition between the self-gra\-vi\-ta\-ting and
not-self-gra\-vi\-ta\-ting regimes at $Q=2$, and the set of parameters $\left(R,
\Sigma\right)$ for which the theoretical viscosities in the two regimes are
equal.}
\label{self_grav_trans}
\end{figure}

As the disks spreads, its surface density will decrease. Transitions from SG to
NSG regimes might thus occur, leading to important variations of the viscosity.
To identify the effect of a self-gravity regime transition on the viscosity, we
study the relative magnitude of the theoretical viscosities in the two regimes,
and analyse the position of the transition $Q = 2$ in the space of
parameters $(R, \Sigma)$. The result is plotted in Fig.
\ref{self_grav_trans}, for particles with $r_{\rm p} = 1\rm~m$.

A transition between the SG to the NSG regime will occur at a given distance R
when the surface density $\Sigma$ goes below the limit represented by the
dashed-black curve for which $Q=2$. Close to the planet, the transition from
the pink to the dark blue region will lead to an increase of the viscosity,
because in these regions $\nu_{\rm SG} < \nu_{\rm NSG}$. Far from the planet,
the transition from the green to the yellow region will lead to a decrease
of the viscosity, because in these regions $\nu_{\rm SG} > \nu_{\rm NSG}$. The
transitions between the two behaviours occurs at $R \sim 78,000 \rm~km$.

This might be explained by looking into the particle velocity dispersion. Inside
a wake, this velocity is the escape velocity: $\sigma_{\rm r}=\sqrt{G
m_{\rm p}/r_{\rm p}}$. In the absence of wakes, the velocity dispersion is the
difference of orbital speed between two particle distant by $2r_{\rm p}$:
$\sigma_{\rm r} = 2 r_{\rm p} \Omega$. Far from the planet, the velocity in a
wake is greater, while close to the planet the difference of orbital velocity
dominates. Equalizing the two expressions shows a transition around $R \sim
84,000 \rm~km$, which is in good agreement with the transition observed in Fig.
\ref{self_grav_trans}.

\subsection{A simple code}
\subsubsection{Numerical procedure}
We have developed a 1D finite elements code on a staggered mesh: the surface
density is estimated on a regular grid, and its evolution is computed by
estimating the mass fluxes at the inner and outer edges of each bin of the grid.
This allows a second order accuracy in space derivatives.

By integrating Eq. (\ref{main_equ}) over the bin width dR, the variation over
a time step dt of the surface density in the bin i is given by:
\begin{equation}
d\Sigma_i = \left[ \frac{\mbox{Flux}\left(R_i + \frac{dR}{2}\right)
- \mbox{Flux}\left(R_i-\frac{dR}{2}\right)}{\rm Bin~Surface~Area} \right]
dt,
\end{equation}
The flux at time t reads:
\begin{equation}
\mbox{Flux}(R) = 6 \pi \left[ \sqrt{R} \frac{\partial}{\partial R} \left( \nu
\Sigma \sqrt{R} \right) \right].
\end{equation}
It is estimated using a second-order (in space and time) explicit Runge-Kutta
scheme.

As in any numerical procedure we can't study scales smaller than the grid
resolution ($\sim 100\rm~km$ per bin). As a result we don't expect to see
small-scale instabilities such as self-gravity wakes, or viscous overstabilities
\citep{latt09}, appear in our simulations. Moreover, as we use a constant value
for the velocity dispersion, we do not satisfy the conditions for viscous
instabilities to appear. Anyway the purpose of this work is to study the rings
evolution on large scales, and even though the full resolution of the
hydrodynamics equations in 2 or 3D would be desirable \citep[e.g. like
in][]{baru08, mass09}, such codes are limited to short time-scales. We keep the
main physical ingredients in this 1D model, allowing us to perform simulations
over 5 billion years.

\subsubsection{Validation}
To validate numerically the behaviour of our code we check that the system's
mass is numerically conserved. We perform a simulation starting with an
initially narrow ring with Gaussian shape and an arbitrary initial mass. The
boundary conditions are set so that material cannot pass through the limits of
the domain: the disk is ``trapped'' inside the domain. We let the ring evolve
viscously and compute the total mass of the disk using the output surface
density. The initial viscous time-scale $\Delta R^2/\nu$ is
$\sim11,500$ years for $\Delta R = 3000\rm~km$. After 100 million years of
evolution (about 9000 viscous time-scales), the mass is equal to its original
value down to the machine precision ($10^{-18}$). Our code numerically conserves
the system's mass.

Equation \ref{main_equ} includes the equation of angular momentum conservation,
so by construction our code should also conserves angular momentum, as long as
the disk has not reached the domain's limits where the boundary conditions cause
modification of the disk's angular momentum. During the period where the disk
spreads freely between the domain's limits, we have checked that angular
momentum is also numerically conserved down to machine precision.

In order to validate the general behaviour of the code, we investigated the
solution given by \cite{prin81} for the viscous spreading of an accretion disk
with constant viscosity. Our numerical results are in good agreement with the
analytical solution, even though slightly different because the width of the
radial grid does not allow to define an initial ring with a perfect Dirac
distribution.

\section{Viscous spreading with variable viscosity}
Here we study the viscous evolution of a ``standard model'' using CV and NCV
models. We present first the disk initial conditions, then the viscous evolution
over 5 billions years, and finally we focus on the evolution of the disk's mass.

\subsection{Description of the standard model}

\subsubsection{Initial profile}
The initial state of Saturn's rings is dimly constrained, as it depends on the
formation scenario, which is still subject to discussions. For instance the
scenario based on the destruction of a satellite via a cometary impact implies
that the satellite that created the rings must be outside the synchronous orbit
and inside the Roche limit for ice at the time of the impact \citep{char09a}.
But we have no further constraints on the initial rings width, position, surface
mass density, particle size distribution\dots~For the sake of simplicity and
generality, the initial conditions are defined as follows:

\paragraph{Initial surface density profile}
We consider at $t=0$ that the disk is a narrow ring $\sim 3000\rm~km$ wide with
Gaussian profile. This is useful to study its evolution over several viscous
time-scales. The ring centre is set slightly outside the synchronous orbit
$\left(\sim 110,000\rm~km\right)$, which is more or less the middle of today's
ring system. The influence of the initial profile is studied in section 4.

\paragraph{Particle radius}
The particle radius of today's rings ranges from a few centimetres to several
metres. There are also radial variations, and on the same orbit different sizes
of particles may coexist \citep{cuzz78, cuzz80, porc08}. Particle sizes should
also evolve through collisions \citep{albe06}. Unfortunately the viscosity model
developed by \cite{dais01} does not include a particle size distribution, but a
fixed particle radius $r_{\rm p}$. \cite{maro83} showed that particles in
Saturn's rings follow a power law $N(r_{\rm p}) \propto r_{\rm p}^{-\alpha}$
with $\alpha \approx 3$ for $R_1 < r_{\rm p} < R_2$. \cite{shu85} then computed
that this distribution can be represented by a single equivalent particle radius
$R_e$ given by $R_e = \sqrt{3}R_2/\pi$. For $R_2 \sim 10\rm~m$ we get $R_e \sim
5\rm~m$. \cite{gold82} suggests an equivalent radius $r_{\rm p} = 1\rm~m$,
which is also a common value in many N-body simulations. Considering the
uncertainty around an equivalent particle size, we choose $r_{\rm p} = 1\rm~m$
for our standard model, and discuss the effect of particle size in section 4.

\paragraph{Initial disk mass}
Considering that today's rings mass is estimated to be about Mimas' mass
\citep{espo83}, one can assume that the initial disk mass was \textit{at least}
Mimas' mass. But on the other hand meteoritic bombardment is believed to bring a
large amount of mass into the rings \citep{cuzz98}, which could substantially
modify the rings' mass. For these reasons we use in our standard model an
initial mass equal to Mimas' mass, and study the initial mass influence in
section 3.3.

\begin{figure*}[!ht]
\begin{center}
\includegraphics[width=14.5cm]{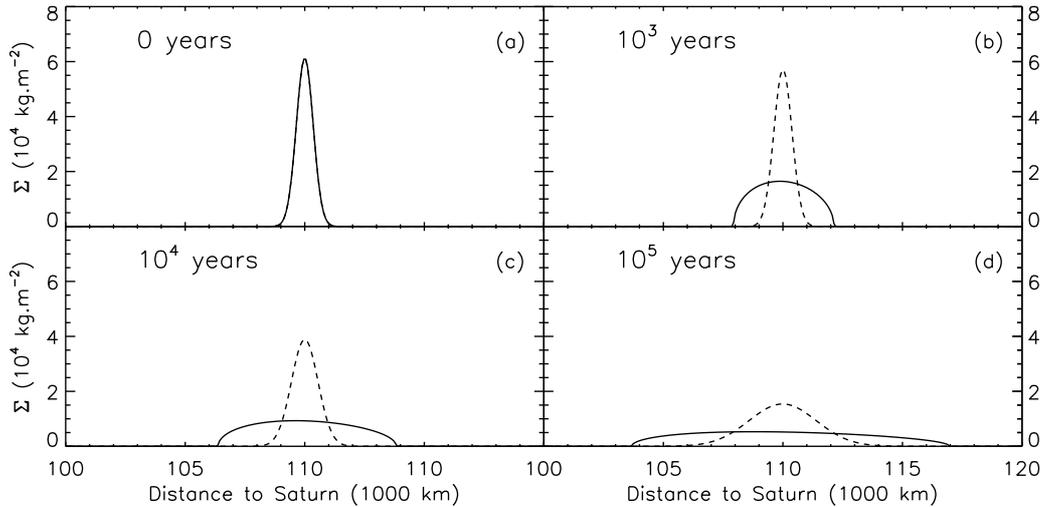}
\end{center}
\caption{Disk surface density at different evolution times for variable
(solid line) and constant (dashed line) viscosities. \textbf{(a)}~Initial
profile. \textbf{(b)}~At 1000 years of evolution. \textbf{(c)}~At $10^4$ years
of evolution. \textbf{(d)}~At $10^5$ years of evolution. The disk with variable
viscosity spreads faster and does not keep the original shape of the density
profile.}
\label{early_evolution}
\end{figure*}

\subsubsection{Constant viscosity values}
To analyse our results with the NCV model, we also perform simulations with a
constant and uniform viscosity model. Assuming that the rings were denser in the
past, which would increase their viscosity, we choose a value of
$0.1\rm~m^2.s^{-1}$, which is about ten times the viscosity of today's A ring
\citep{tisc07}. Anyway, in the CV case, the value of the viscosity does not
modify the density profile, but only the viscous time-scales.

\subsubsection{Boundary conditions}
The radial domain ranges from $R=65,000\rm~km$ to $R=140,000\rm~km$ ($R=0$
is the centre of Saturn). The boundary conditions are set as follows: material
is released from the disk when it passes the domain's limits. For the inner
boundary we consider that the material falls onto the planet and is destroyed,
and for the outer boundary we consider that the material crosses the Roche limit
and starts to be accreted in small moons that do not interact viscously with the
disk. They should however interact with the disk at resonances, but this is
beyond the scope of this paper. This has been investigated by \cite{char10}
(\textit{In press}).

We remove the material passing through the domain's limits, and the
corresponding angular momentum, by setting the surface density of the first and
last bins to zero at $t=0$, and at each time step we equalize the inward and
outward fluxes in these bins. As a result, the total flux on the first and last
bins is 0, and all material entering these bins is immediately removed, along
with the corresponding angular momentum. No torque is exerted on the ring by the
boundary.

\subsection{Viscous spreading over 5 billion years}
In this section we study the evolution of the standard model, using constant and
variable viscosities, over the age of the Solar System $\left(\sim
5\rm~billion~years\right)$.

\subsubsection{Relevant parameters}
To analyse our results we study the evolution of three quantities: 
\begin{itemize}
\item the surface density profile;
\item the total mass of the disk, derived from the surface density, in order to
track the mass lost during the viscous spreading;
\item the total mass passing through the left and right limits of the domain.
\end{itemize}

\subsubsection{Early evolution ($0$ - $10^5$ years)}
In Fig. \ref{early_evolution} the surface density at different times for both CV
and NCV models is plotted: at 0 and after $10^3$, $10^4$ and $10^5$ years of
evolution. As in \textbf{section 2.3.2}, the initial spreading time-scale is
$\rm \sim 11,500\rm~years$.

In the CV case, the disk edges remain smooth during the spreading, while steep
edges are created in the NCV case (Fig. \ref{early_evolution}b). In this latter
case, $\nu$ being an increasing function of $\Sigma$, the viscosity is high in
the centre and low at the edges. Thus the centre spreads faster than the edges.
As a result, a lot of material is transported from the core towards the edges,
but the edge itself fails to spread this great amount of material and is
``overwhelmed'' by material coming from the core (Fig. \ref{early_evolution}b,
solid line). As a result, the surface density rapidly increases close to the
edges, which become steep.

\begin{figure*}[!ht]
\begin{center}
\includegraphics[width=14.5cm]{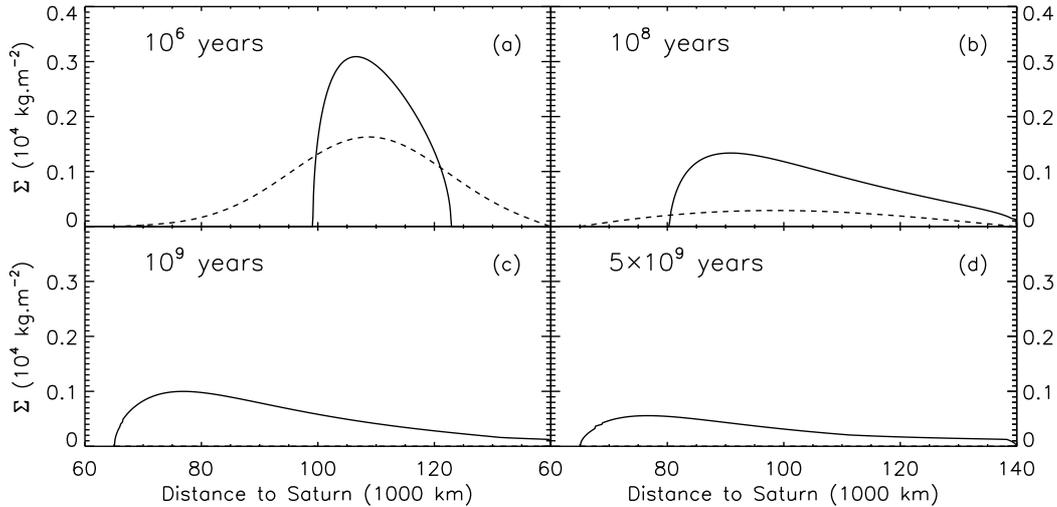}
\end{center}
\caption{Disk surface density at different evolution times with
variable (solid line) and constant (dashed line) viscosities.
\textbf{(a)}~At 1Myrs of evolution. \textbf{(b)}~At 100 Myrs of evolution.
\textbf{(c)}~At 1 Gyrs of evolution. \textbf{(d)}~At 5 Gyrs of evolution. The
disk with constant viscosity is emptied in $\sim 10^9$ years, while the disk
with variable viscosity remains massive over 5 Gyrs with a density peak inward
and lower densities outward.}
\label{late_evolution}
\end{figure*}

At the beginning the NCV disk appears to spread faster than the CV disk: after
$10^5$ years of evolution (Fig. \ref{early_evolution}d), the disk width is about
$1.5 \times 10^4$ km in the NCV case, and $10^4$ km in the CV case. This can be
explained by the fact that the initial value for the CV model is too small. The
computed viscosity at the rings centre is initially $\sim100\rm~m^2.s^{-1}$,
much larger than the $0.1\rm~m^2.s^{-1}$ value we use for the CV model. But
after $10^5$ years of evolution the viscosity drops to $\sim 0.8\rm~m^2.s^{-1}$.
Thus whereas the initial spreading of the NCV disk is very rapid due to high
viscosity, as soon as the disk has spread, the viscosity drops and the evolution
slows down dramatically.

\subsubsection{Evolution over 5 billion years}
In Fig. \ref{late_evolution} the surface mass density at  $10^6$, $10^8$, $10^9$
and $5\times10^9$ years is plotted.

We see here that the CV disk has spread much faster than the NCV disk (Fig.
\ref{late_evolution}a). After 1 billion years, the CV disk is no longer visible
because almost all material has been spread out of the rings (Fig.
\ref{late_evolution} (c) and (d)). Its average surface density is $~ 5\times
10^{-5}\rm~kg.m^{-2}$. The NCV disk has evolved significantly, and memory of
the Gaussian initial conditions has been completely lost.

The NCV disk evolves only very little from 1 to 5 billion years. This is due to
a drop of the viscosity because of the surface mass density being now much lower
than initially, resulting in a much slower evolution. After 1 billion years of
evolution, the rings viscosity is $\sim 10^{-3}\rm~m^2.s^{-1}$.

In conclusion, this study illustrates that spreading time-scales estimated using
only $\Delta R^2/\nu$ with a constant viscosity \citep{gold82, espo86}, are
quite inaccurate when compared with our numerical simulations that include a
realistic viscosity model. As a result the viscous spreading rate of the rings
is far from being linear.

\subsubsection{Evolution of the disk's edges}
Initially, due to the Gaussian shape we use, the disk's edges are smooth. Due to
a low surface density, the region close to the edges is initially in the
non-self-gra\-vi\-ta\-ting regime. When the disk starts to spread, material is
transported from the core toward the less viscous edges. This increases the
surface density close to the edges, which then become self-gra\-vi\-ta\-ting. As
the disk continues to spread, the surface density close to the edges decreases.
Eventually, the edges become non-self-gra\-vi\-ta\-ting again. 

\begin{figure}[!h]
\begin{center}
\includegraphics[width=8.5cm]{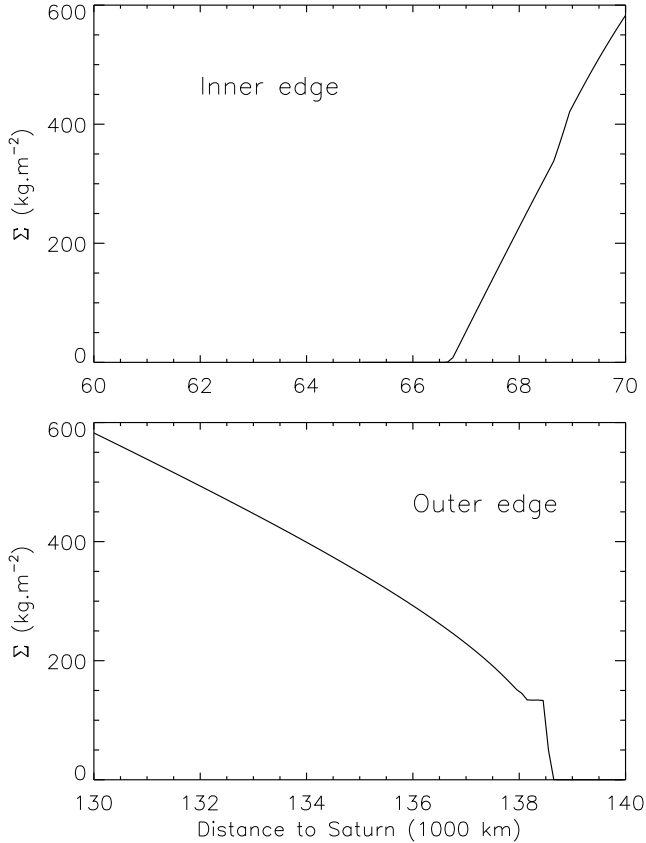}
\end{center}
\caption{Zoom on the disk's edges close to the domain's limits. The
inner edge is smooth and progressive (Top), while the outer edge is sharp
(bottom).}
\label{disk_edges}
\end{figure}

As mentioned in section 2, the transitions from SG to NSG areas should lead to
changes in the spreading of the disk, with different behaviours depending on the
distance to the planet. Close to the planet, this leads to an increase of the
viscosity (see Fig. \ref{visco_trans_coll} black dashed and dotted
lines), which becomes inversely proportional to R: the closer to the planet,
the higher the viscosity (as it is in this case dominated by the collisional
component). As a result, the inner edge is smoothed (Fig. \ref{disk_edges},
top).

Far from the planet, the opposite process occurs. Here, the transition from SG
to NSG results in the viscosity dropping by a factor $\sim100$. The material
flux is proportional to $\nabla (\nu\Sigma)$ (Eq. (\ref{main_equ})). When
the viscosity is constant, this can be rewritten as $\nu \nabla \Sigma$, so
that the flux depends only on the surface density gradient. With variable
viscosity, this simplification is no longer valid: the flux depends on the
gradient of both the viscosity and the surface density: $\nabla (\nu\Sigma)
= \left(\nabla\nu\right)\Sigma + \nu\left(\nabla\Sigma\right)$.

Firstly, in the NSG regime, the viscosity dependence on R is $\nu_{\rm trans}
\propto R^{3/2}$ and $\nu_{\rm coll} \propto R^{-3/2}$, while in the SG regime
$\nu \approx \nu_{\rm trans} \propto R^{19/2}$ (Eqs \ref{nu_trans_eq} and
\ref{nu_coll_eq}). As a result, the viscosity gradient is much lower in the NSG
regime than in the SG regime. Secondly, due to the presence of a steep edge, the
density gradient is larger close to the edge than in the core. However, when the
edge enters the NSG regime, the viscosity significantly decreases and damps the
large surface density gradient. Globally, the material flux is greater close to
the outer edge than \textit{at} the edge, which causes material to accumulates
and creates the observed plateau  (Fig. \ref{disk_edges}, bottom). 

\begin{figure}
\begin{center}
\includegraphics[width=8.5cm]{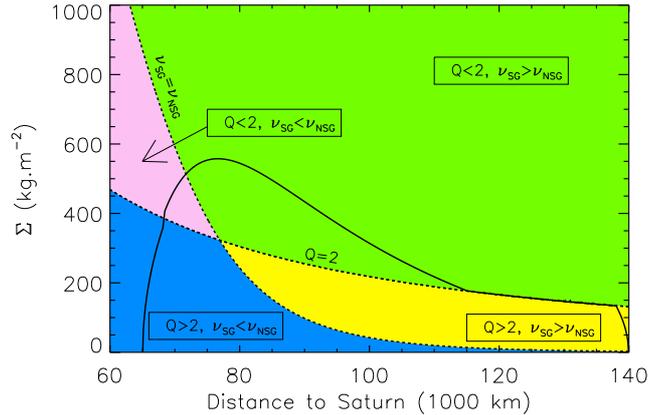}
\end{center}
\caption{Disk's surface density at $t=5\rm~Gyrs$ (solid black curve),
and self-gra\-vi\-ta\-ting regime. Regions of the disk for which the surface
density is above the dashed-curve ``$Q=2$'' are self-gra\-vi\-ta\-ting, while
those below are not-self-gra\-vi\-ta\-ting.}
\label{sg_trans_surf_dens}
\end{figure}

These opposite behaviours can be more appreciated using the
self-gravity-regime-transition diagram presented in section 2.2.2. In Fig.
\ref{sg_trans_surf_dens} the disk's surface density at $t=1\rm~Gyrs$ is plotted
over the map of the different viscosity regimes as a function of R and $\Sigma$.
It shows in particular that the outer plateau closely follows the limit $Q=2$.
Note that this marginally gravitationally stable plateau is similar to Saturn's
A ring, which is known to be slightly gravitationally unstable, with the regular
appearance and disappearance of self-gravity wakes \citep{ferr09}.

\subsubsection{Spreading time-scales}
\begin{figure}[!h]
\begin{center}
\includegraphics[width=8.5cm]{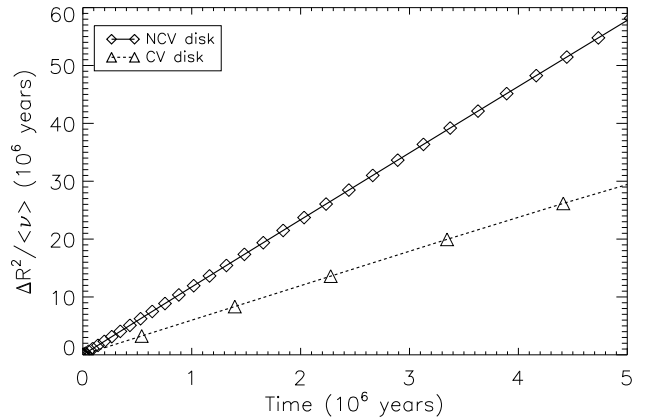}
\end{center}
\caption{Evolution of the ratio disk's width over viscosity, against
time, for a disk with constant (dashed line) and variable (solid line)
viscosity. The diamonds and triangles are data points from simulations. Note how
$\Delta R^2/\nu$ remains proportional to time.}
\label{disk_width_viscosity_evolution}
\end{figure}

We look for semi-analytical solutions of Eq. (\ref{main_equ}) to address the
difference of spreading time-scales between the disk with constant and variable
viscosities. In a classical mass diffusion process the standard deviation of the
mass distribution $\Delta R^2$, increases linearly with time, so that $
\Delta R^2 / \nu$ is proportional to time. In Fig.
\ref{disk_width_viscosity_evolution} we plot this quantity for the CV and NCV
disks. In order to avoid any boundary effect, we restrict the analysis to the
$t < 5\times10^6\rm~years$ where the disk has not yet reached the domain's
limits.

The  width of the ring is defined as 
\begin{equation}
\Delta R^2 = \frac{\sum_i \left( R_i - R_{\rm c} \right)^2
\Sigma_i}{\sum_i \Sigma_i},
\end{equation}
where $R_{\rm c}$ is the ``centre'' of the disk defined by 
\begin{equation}
R_{\rm c} = \frac{\sum_i R_i\Sigma_i}{\sum_i
\Sigma_i},
\end{equation}
For the disk with variable viscosity, we have used the mean viscosity, weighted
by the bin's mass
\begin{equation}
<\nu> = \frac{\sum_i M_i \nu_i}{\sum_i M_i},
\end{equation}
where $M_i = 2\pi R_i dR\Sigma_i$ is the bin's mass and dR is the bin's
width.

It is remarkable that despite the change of viscosity by several orders of
magnitude during the evolution of the system, $\Delta R^2 / \nu$ remains
closely proportional to time at any instant (Fig.
\ref{disk_width_viscosity_evolution}). However, this expression can no longer be
used as a ``rule of thumb'' to estimate, with the initial disk's parameters, the
spreading time-scales as the viscosity drops sharply during the ring's
evolution.

\begin{figure}[!h]
\begin{center}
\includegraphics[width=8.5cm]{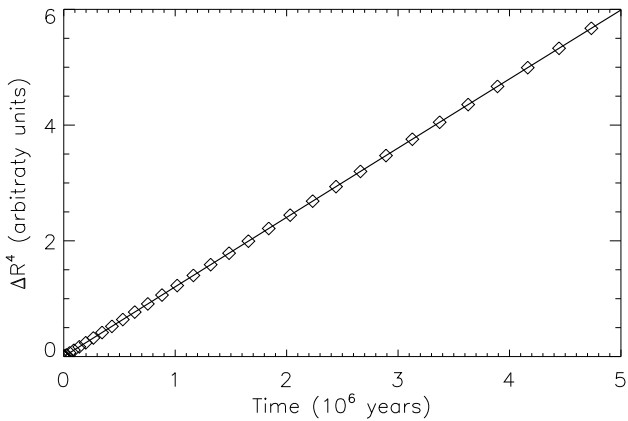}
\end{center}
\caption{Disk width evolution against time with variable viscosity. The
diamonds are data points from simulations. Note how $\Delta R^4$ remains
proportional to time, so that the disk's width scales like $t^{1/4}$.}
\label{disk_width_evolution}
\end{figure}

For a disk in the self-gra\-vi\-ta\-ting regime, the viscosity is dominated by
the translational and gravitational components, and can be expressed as $\nu
\propto \Sigma^2$. In a rough description we assume that the disk is a slab of
material of width L and centred on $R = R_0$. In that case, the surface
density can be expressed as $\Sigma = M_0/\left( 2 \pi R_0 L\right)$, where
$M_0$ is the disk's initial mass. Using $L^2/\nu \propto t$ we can then
write $L^4 \propto t$. To check the validity of this relation, the disk's
width evolution against time is plotted in Fig. \ref{disk_width_evolution}. We
obtain a perfect proportionality relation.

For a disk with constant viscosity, we directly get $L^2 \propto t$.
Inverting this relation we get that the width of a disk with constant viscosity
increases as $t^{1/2}$, while with variable viscosity it increases as $
t^{1/4}$. The spreading time-scale of a disk with variable viscosity is then
much smaller than that of a disk with constant viscosity.

\subsection{Mass evolution}
Once the disk reaches the domain's limits (either inner or outer), it empties
out. We describe below the emptying time-scale and show that it is strongly
dependant on the disk's initial mass and self-gra\-vi\-ta\-ting regime.
\subsubsection{Disk mass}
The total mass of the disk is given by: 
\begin{equation}
M_{\rm disk} = \int_0^{\infty} \Sigma \left( R \right) 2 \pi R dR
\label{mass_compute}
\end{equation}
On a discrete grid, this equation becomes
\begin{equation}
M_{\rm disk} = \sum_{i=0}^{N-1} \Sigma_i \pi \left( R_{i+1}^2 - R_i^2
\right)
\end{equation}
where N is the number of bins in the grid.

\begin{figure}[!h]
\begin{center}
\includegraphics[width=8.5cm]{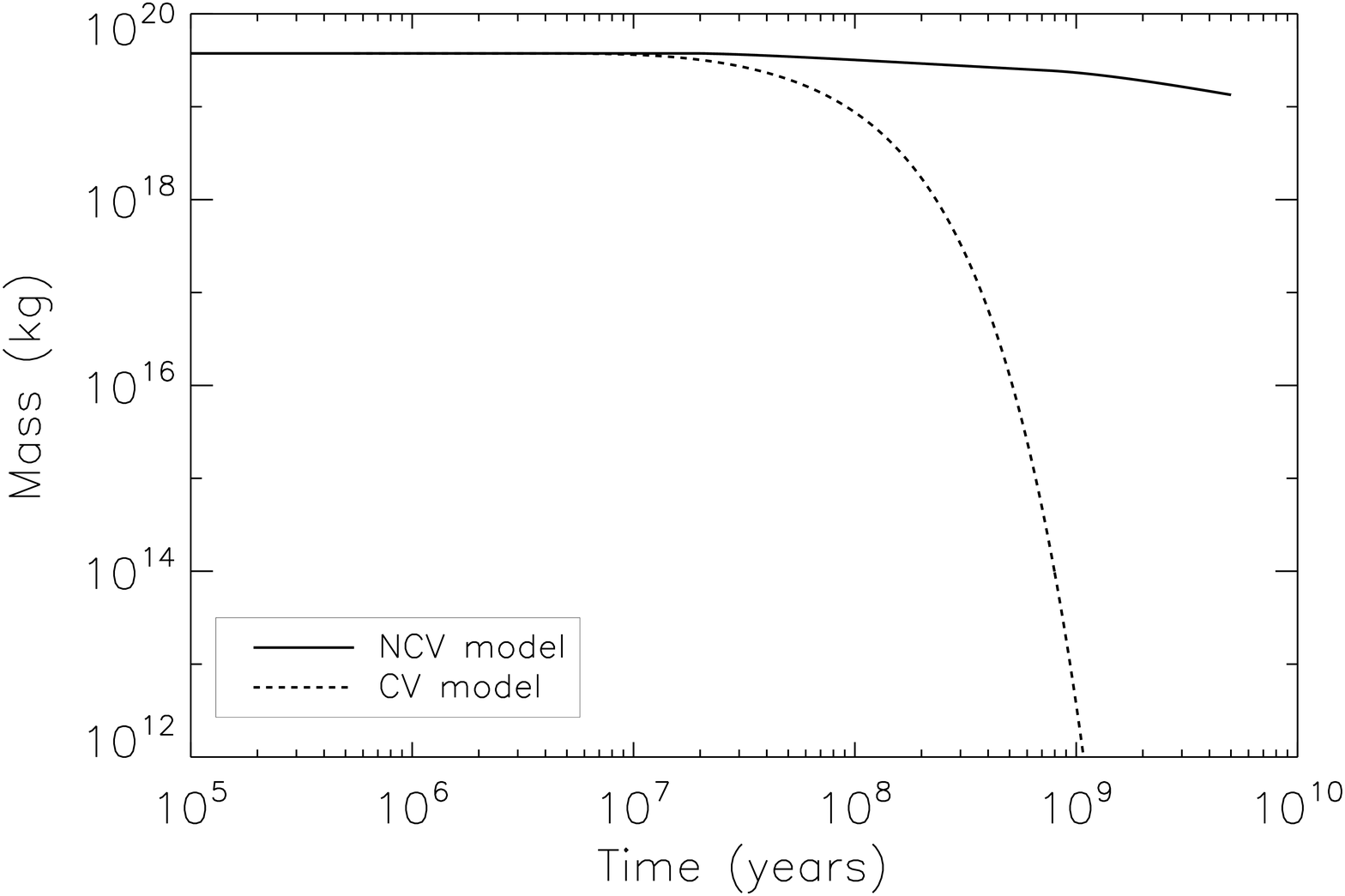}
\end{center}
\caption{Disk mass as a function of time, for a disk with variable
(solid line) and constant (dashed line) viscosity. The disk with constant
viscosity is emptied in $\sim 10^9\rm~years$, while a third of the initial mass
remains at 5 Gyrs in the disk with variable viscosity.}
\label{mass_evolution}
\end{figure}

In Fig. \ref{mass_evolution} is plotted the evolution of the disk's mass over 5
billion years, with the CV and NCV models. The NCV case can be divided into 3
successive steps. First the disk's mass remains constant as no material has yet
reached the domain's limits. Second, at $2\times10^7$ years the disk reaches the
domain's outer limit and its mass starts to decrease. As seen in the previous
section, the disk spreads with a lot of material being stacked close to the
stiff outer edge. As a result, a lot of mass is rapidly lost by the disk when it
reaches the domain's outer limit. Third, at $\sim10^9$ years, the disk reaches
the domain's inner limit and the mass loss is increased.

While the disk is completely emptied in a few 100 million years in the CV case,
the disk mass is still $\sim1 \times 10^{19}\rm~kg$ after 5 billion years in the
NCV case. It seems then possible for planetary rings to survive viscously over
the age of the Solar System in the dynamical environment of Saturn when
considering a physically realistic viscosity model.

\subsubsection{Initial mass influence}
One can argue that the survival of $\sim10^{19}\rm~kg$ of material at the end of
the disk's evolution is only valid for the set of initial parameters considered
here, in particular the initial mass of $3.75\times10^{19}\rm~kg$. With a lower
initial mass, there would be less material to evacuate from the disk. But it
would also decrease the surface density and the viscosity, and thus would slow
down the loss of material. The opposite behaviour would occur with a higher
initial mass. Then it raises the question: how does the disk's final mass
depend on the initial mass ?

\begin{figure}[!h]
\begin{center}
\includegraphics[width=8.5cm]{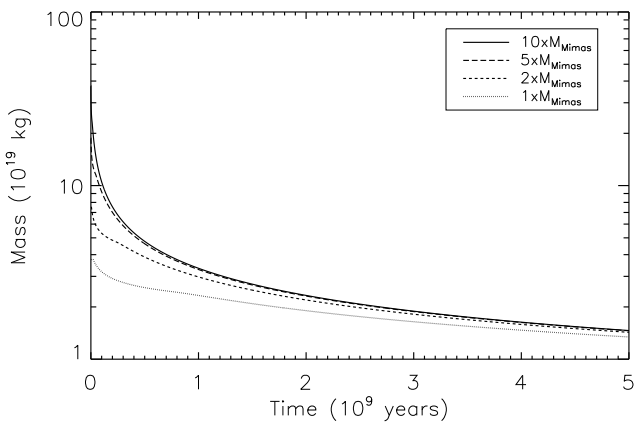}
\end{center}
\caption{Disk mass as a function of time, for different disk initial
masses. With a higher mass, the loss-of-mass rate is increased initially but
drops down over time, so that all disk's mass are comparable after 5 Gyrs of
evolution.}
\label{mass_init_influence}
\end{figure}

Today's mass of Saturn's rings have been evaluated to be similar to Mimas' mass,
or probably more \citep{espo83}. Even if meteoritic flux is expected to bring a
significant amount of material to the disk \citep{cuzz98}, it seems natural to
assume that the rings initial mass was larger than their present mass. Anyway,
to address this question we performed a series of simulations starting with
initial masses ranging from 1 to 10 Mimas masses. Results are plotted in Fig.
\ref{mass_init_influence}.

The disk that starts with 10 Mimas masses (i.e. $3.75\times 10^{20}\rm~kg$ )
loses
$3\times 10^{20}\rm~kg$ in the first 100 million years: that's 80 \% of its
initial mass ! So the initial mass of the disk does modify the spreading speed
and the amount of material ejected: adding more mass to the initial disk
enhances the viscosity and a large amount of material is then ejected from the
disk very rapidly. In particular, it is interesting to note that in order to
form rings with a mass close to Mimas' mass (i.e. $\sim3\times 10^{19}\rm~kg$),
they should not be older than 1 billion years old, starting from 10 Mimas
masses.

On a larger time-scale, the loss rate drops. Interestingly, the emptying rate of
all disks slows down significantly during the ring's evolution. After 4 billion
years, no further significant evolution is seen, and at 5 billion years, all
disk masses are $\sim1.5\times10^{19}\rm~kg$.

\subsubsection{Analytical insight}
We investigate whether asymptotic laws can be used to describe the evolution of
the mass lost by the disk, in the self-gra\-vi\-ta\-ting and
non-self-gra\-vi\-ta\-ting
regime. 

\paragraph{Self-gra\-vi\-ta\-ting regime}
Initially the disk is fully self-gra\-vi\-ta\-ting. As showed in section 3.2.5
we can
then write that $L^2 \propto \left(M_0/\left(2\pi R_0L\right)\right)^2t$.
Finally we get $L^4 \propto M_0^2t$.

We can use the above law to estimate the mass lost by the disk. Assuming that
the disk fills the entire area available for its spreading, we get L = constant
and the further ``widening'' of the rings is equivalent to a loss of mass,
because all material passing through the domain's boundaries is removed from the
disk. Using the above expression we can express a time-scale for the emptying of
the disk $t_{\rm empty} \propto 1/(M_0^2) $.

\begin{figure}[!h]
\begin{center}
\includegraphics[width=8.5cm]{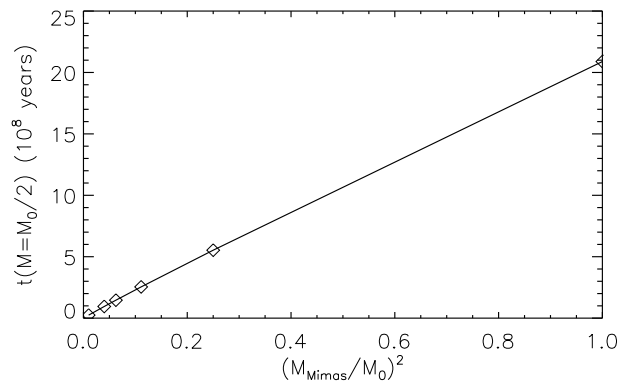}
\end{center}
\caption{Timescale for a 2-fold decrease of the disk mass. The
diamonds are data points from simulations. Note that it is proportional to
$1/M_0^2$, where $M_0$ is the initial disk mass.}
\label{empty_time}
\end{figure}

To compare this law with our simulations, the time needed for the disk's mass to
be half the initial mass, against $1/M_0^2$, is plotted in Fig.
\ref{empty_time} for initial masses of 1, 2, 5 and 10 masses of Mimas. We obtain
a relation of proportionality that confirms the above analytical study: in the
self-gravity regime, the disk is emptied on a time-scale inversely proportional
to the square of its initial mass. Is this relation still valid for a disk in
the non-self-gra\-vi\-ta\-ting regime ?

\paragraph{Non-self-gra\-vi\-ta\-ting regime}
In the NSG regime, the gravitational viscosity is 0 and the collisional and
translational viscosities have comparable magnitudes (see Fig.
\ref{visco_trans_coll}). Their dependence on $\Sigma$ is $\nu_{\rm coll} =
r_{\rm p}^2\Omega \beta \Sigma$ with $\beta = 3/(4 r_{\rm p} \rho_{\rm p})$,
and $
\nu_{\rm trans} = \gamma \beta \Sigma / (1 + \beta^ 2 \Sigma^ 2)$ with $\gamma =
0.46 \sigma_{\rm r}/(2 \Omega)$. The total viscosity is then 
\begin{equation}
\nu = \nu_{\rm coll} + \nu_{\rm trans} = \frac{A \Sigma + C \Sigma^3}{1 +
\beta^2
\Sigma^2},
\end{equation}
where $A = \beta \left( r_{\rm p}^2 \Omega + \gamma \right)$ and $C = r_{\rm
p}^2
\Omega \beta^3$. Using the same simple approximation $\Sigma = M_0/(2 \pi
R_0 L)$ and using $L^2/\nu = Kt$ (K is a proportionality constant) we get 
\begin{equation}
\frac{L^5 + a L}{c(b^2 L^2 +1)} = Kt,
\label{equ_L5aL}
\end{equation}
where $a=(\beta M_0/(2 \pi R_0))^2$, $b = \sqrt{A/C}(2 \pi R_0/M_0)^2$
and $c = C(M_0/(2 \pi R_0))^3$.

For standard parameters of Saturn's rings ($L \approx 10^4\rm~km$, $R_0
\approx 10^5\rm~km$, $r_{\rm p}=1\rm~m$, $\rho_{\rm p}=900\rm~kg.m^{-3}$) we
get $L^5 \approx 10^{35}$, $aL \approx 10^{21}$, $b^2L^2 \approx 10^2$. We can
then simplify Eq. (\ref{equ_L5aL}) as:
\begin{equation}
Kt \approx \frac{L^5}{cb^2 L^2}.
\label{equ_L5}
\end{equation}
Finally the emptying time-scale verifies $t_{\rm empty} \propto 1/M_0 \gg
1/M_0^2$. As a result the emptying time-scale is much larger in the NSG regime
than in the SG regime. In the following section we investigate the transitions
between these two regimes, in terms of disk's mass.

\subsubsection{Mass of a fully non self-gra\-vi\-ta\-ting disk}
As the disk spreads, the surface density drops down and larger parts of the disk
(near the edges) become marginally self-gra\-vi\-ta\-ting. We should then see
the
emptying time-scale drop significantly over time. Fig. \ref{mass_init_influence}
shows indeed an initial rapid emptying of the disk, which then considerably
slows down. All disk's mass even seem to converge toward a limiting mass. 

We have shown in the previous paragraph that the emptying time-scale is much
increased when the disk is in the NSG regime. In order to find the transition
between the SG and NSG regimes, we look for a condition on the disk's mass for a
disk to be self-gra\-vi\-ta\-ting or not. We investigate this condition in the
case
where the disk fills the entire area available for its spreading (that is from
the planet's radius to the Roche limit). We then use the Toomre's Q parameter to
describe the self-gra\-vi\-ta\-ting regime of the disk.

This parameter is a measure of the gravitational stability of the disk. As
mentioned earlier we consider here that the transition between the SG and NSG
regimes occurs for $Q = 2$. The disk is then fully non-self-gra\-vi\-ta\-ting if
:
\begin{equation}
\forall R, Q=\frac{\Omega \sigma_{\rm r}}{3.36 G \Sigma} > 2.
\label{Qapprox2}
\end{equation}
Using $\Omega = \sqrt{G M/R^3}$ and $\sigma_{\rm r} = \sqrt{G m_{\rm p}/r_{\rm
p}}$ we
find that the disk is completely NSG if the surface density verifies 
\begin{equation}
\forall R,  \Sigma < 0.15\sqrt{\frac{M m_{\rm p}}{r_{\rm p}}} R^{-3/2}.
\label{Qapprox2_2}
\end{equation}
Then using  (\ref{mass_compute}) and for the specific case of Saturn, we find
that the disk is completely in the non-self-gra\-vi\-ta\-ting regime if its
total mass
verifies
\begin{equation}
M_{\rm disk} < M_{\rm NSG} \approx 1.032\times 10^{19}~\rm kg,
\end{equation}
In the simulations presented in this section, the disk's mass at t=5 Gyrs is
$\sim1.34-1.46\times10^{19}~\rm kg$ for initial masses of 1 to 10 Mimas' masses.
Theses masses are larger than $M_{\rm NSG}$ which should indicate that they are
still in the self-gra\-vi\-ta\-ting regime (at least partially). This is indeed
the
case as seen in Fig. \ref{sg_trans_surf_dens}

In conclusion, it seems that whatever the initial mass, the disk undergoes an
initial rapid evolution when it is self-gra\-vi\-ta\-ting. This evolution
continuously
slows down as larger parts of the disk become marginally self-gra\-vi\-ta\-ting.
Eventually, the disk will becomes entirely NSG, and will undergo a very slow
evolution. The wakes observed in Saturn's A ring \citep{ferr09} indicate that
the ring is still self-gra\-vi\-ta\-ting, at least in some regions.

\subsubsection{Mass fluxes through boundaries}
In this section, we consider again the previous standard model with constant and
variable viscosities, starting with 1 mass of Mimas. We investigate the mass
lost by the disk through the inner and outer domain's limits, and whether theses
two quantities are comparable or not. This can give clues on the mass falling
onto the planet or crossing the Roche zone, the latter being for instance
available for accretion into satellites \citep[][\textit{In press}]{char10}.

\begin{figure}[!h]
\includegraphics[width=8.5cm]{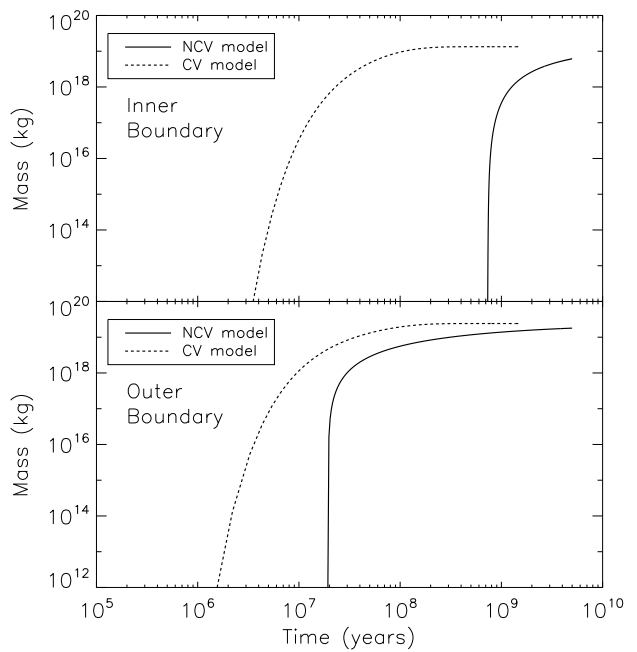}
\caption{Mass passing through domain boundaries (cumulative) with
variable (solid line) and constant (dashed line) viscosities. The data for the
disk with constant viscosity is truncated at t=1.5 Gyrs, as the disk was empty
at that time (see Fig. \ref{mass_evolution}). Whatever the viscosity model, the
disk loses much more mass through the Roche limit (bottom graph) than by infall
onto the planet (top graph).}
\label{mass_dump_lr}
\end{figure}

At each time step, the mass fluxes at the limits of the domain are tracked.
Results are plotted in Fig. \ref{mass_dump_lr}. The dashed line, representing
the mass flow for the CV model, is truncated at $\sim1$ billion years of
evolution,
because at that point the disk was almost emptied out of its material (see Fig.
\ref{mass_evolution}), which caused the numerical code to stop.

\begin{figure*}[!ht]
\begin{center}
\includegraphics[width=16cm]{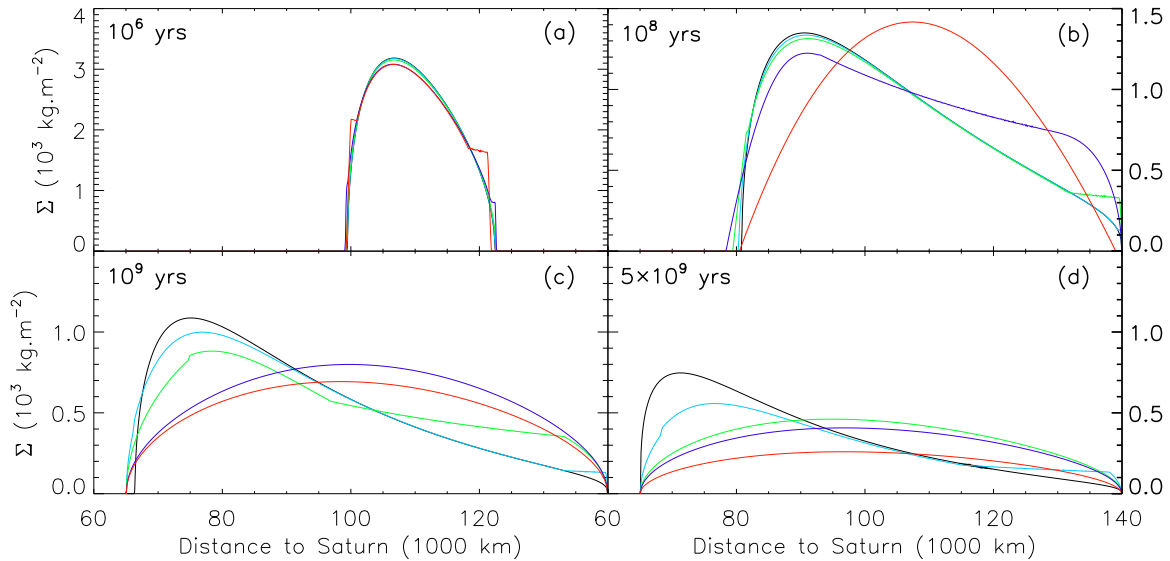}
\end{center}
\caption{Disk surface density at different times and for different
particle radii: 0.01 m (black), 1 m (light blue), 2.5 m (green), 5 m (dark
blue), 10 m (red). \textbf{(a)}  After 1 Myrs of evolution. \textbf{(b)} After
100 Myrs of evolution. \textbf{(c)} After 1 Gyrs of evolution. \textbf{(d)}
After 5 Gyrs of evolution. The disks with larger particles enter more rapidly
the non-self-gra\-vi\-ta\-ting regime. A disk still partially
self-gra\-vi\-ta\-ting has a specific shape with an inward peak and an outer
plateau corresponding to the non-self-gra\-vi\-ta\-ting area.}
\label{rp_influence_surf_dens}
\end{figure*}
\ \par
NCV and CV models behave very differently. For the CV disk, both limits of the
domain are reached at about the same time, $\sim10^ 6\rm~years$. For the NCV
disk, both limits are reached much later and not simultaneously: the outer limit
is reached after 20 million years, while the inner limit is reached only after
800 million years. This is due to the much larger viscosity far from the planet,
as the translational component scales like $R^{5/2}$ in the
self-gra\-vi\-ta\-ting regime.

Another feature, not visible in the log-log format, is the very sudden release
of material through the outer limit (due to the stiff outer edge, see section
3.2.4) compared to a smoother and more progressive release through the inner
edge (due to the smooth inner edge).

\section{Influence of ring parameters}
In order to validate our results of section 3, we now perform additional
simulations, varying the disk parameters. In the following, only the variable
viscosity model (see \textbf{section 2.2}) is considered.

\subsection{Influence of ring particles radius}
The viscosity model, and our 1-dimensional approach requires the fixing of the
value of the particle size. In this section, we study the influence of this
parameter. We perform several simulations of the ``standard model'', with single
particle sizes ranging from 0.01 to 10 metres, and investigate the impact on the
evolution of the surface density and the disk's mass, and on the mass fluxes
through the boundaries.

\subsubsection{Influence on surface density evolution}
For a NSG disk, the viscosity increases with particle radius (see section
2.2.2.). For a SG disk, the translational viscosity does not depend on the
particle radius. However, the particle radius increases Toomre's Q parameter. As
a result the disk is more likely to become NSG with bigger particles in regions
with lower surface densities, i.e. close to the edges. Thus, different spreading
regimes may then be expected whether the disk remains SG during spreading or
not. In Fig. \ref{rp_influence_surf_dens} the surface mass density of the disk
at different times, and for several particle sizes, is plotted. 

At 1 million years (Fig. \ref{rp_influence_surf_dens}a), the surface density is
not significantly affected by particle size, except the disk with 10 m-radius
particles (red curve). The ramps close to the disk extremities are due to
transitions from the SG regime to the NSG regime (see section 3.2.4).

At $10^8$ years (Fig. \ref{rp_influence_surf_dens}b) the disk with 10 m-radius
particles has evolved very differently from the other disks. This disk has a
parabolic shape whereas the disks with smaller particles have a surface density
peak inward and lower densities in the outer regions. This latter shape is
specific of a disk that is still partially self-gra\-vi\-ta\-ting (in the peak
region).

\begin{figure}[!h]
\begin{center}
\includegraphics[width=8.5cm]{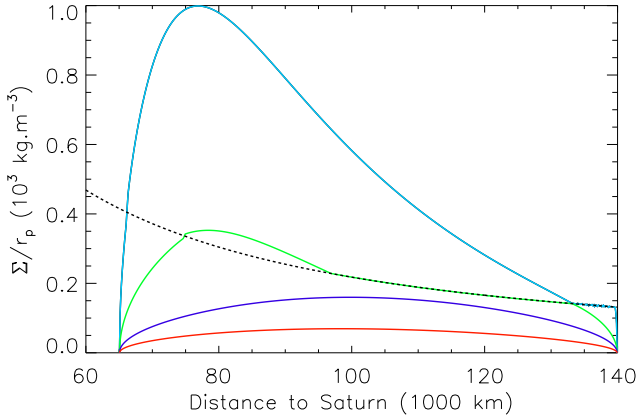}
\end{center}
\caption{Surface density at $t=10^9\rm~years$ normalized by the ring
particle radius for different particle radii: 1 m (light blue), 2.5 m (green),
5 m(dark blue), 10 m (red). The case $r_{\rm p}=0.01\rm~m$ has been remove for
scaling purpose. The black dashed line is the transition at $Q=2$ between the
self-gra\-vi\-ta\-ting to the non-self-gra\-vi\-ta\-ting regime normalized by
the ring particle's radius. The disks with 1m and 2.5 m-radius particles are
still partially in the self-gra\-vi\-ta\-ting regime (in the peak), while the
disks with 5 m and 10 m-radius particles are entirely in the
non-self-gra\-vi\-ta\-ting regime.}
\label{sg_trans_surf_dens_rp}
\end{figure}

At $10^9$ years (Fig. \ref{rp_influence_surf_dens}c), the disk with 5 m-radius
particles is also completely in the NSG regime and its shape is very similar to
the disk with 10 m-radius particles. The disk with 2.5 m-radius particles is in
the SG regime only in the region located between the discontinuity at
$\sim 75,000\rm~km$ and the beginning of the outer ramp at $\sim 95,000\rm~km$.
The surface density of each disk, normalized by the $r_{\rm p}$, is plotted on
Fig. \ref{sg_trans_surf_dens_rp}, along with the transition between the SG and
NSG regimes $\left(Q=2\right)$, also normalized to $r_{\rm p}$ (black dashed
line). 

The dark blue and red curves, corresponding to the disks with 5 m and 10
m-radius particles, are below the dashed line, indicating that they are entirely
no longer in the SG regime. The light blue and green curves, corresponding to
the disks with 1 m and 2.5 m-radius particles, are still partially in the SG
regime, in the regions where the surface density is above the black dashed line.
At $5\times10^9\rm~years$ (Fig. \ref{rp_influence_surf_dens}d), only the disks
with 0.01 m-radius and 1m-radius particles are still partially
self-gra\-vi\-ta\-ting. 

Two spreading regimes clearly appear. In particular, the inner peak of the
surface density is characteristic of a disk that is still partially
self-gra\-vi\-ta\-ting, whereas a parabolic shape is characteristic of a fully
not-self-gra\-vi\-ta\-ting disk.

Increasing the particles size slows down the initial outer spreading, but when
all disks are completely NSG, the disks with the larger particles spread faster,
as viscosities in the NSG regime are an increasing function of $r_{\rm p}$. As a
result, Fig. \ref{rp_influence_surf_dens}d can be interpreted as different
evolutionary steps in the disk's life.

\subsubsection{Influence on disk mass evolution}
The final disk mass for all particle radii is indicated in Table
\ref{table_rp_mass}.
\begin{table}[!h]
\caption{Influence of particle radius on the evolution of the disk's mass
evolution. Initial mass is $3.75\times10^{19}~\mbox{kg}$}.
\begin{tabular}{l l l l}
\hline
Particle & Final & Mass through & Mass through\\
radius & disk mass  & inner edge & outer edge\\
(m) & $\left(10^{19}\rm~kg\right)$ & $\left(10^{19}\rm~kg\right)$ &
$\left(10^{19}\rm~kg\right)$\\
\hline
0.01 & 1.45 & 0.46 & 1.84\\
1 & 1.34 & 0.61 & 1.80\\
2.5 & 1.63 & 0.61 & 1.51\\
5 & 1.46 & 0.69 & 1.60\\
10 & 0.93 & 0.93 & 1.89\\
\hline
\end{tabular}
\label{table_rp_mass}
\end{table}
The final mass of the disk is not much affected by the change of particles size,
except the disk with 10 m-radius particles because of the strong enhancement of
the translational and collisional viscosities in the NSG regime.  The disk with
2.5 m-radius particles is intermediate between a disk whose viscosity remains
high because it remains in the SG regime (for $r_{\rm p} \lesssim 1\rm~m$), and
one that rapidly enters the NSG regime, but whose viscosity is supported by
large particles (for $r_{\rm p} \gtrsim 5\rm~m$).

The inner mass flux increases with the particle radius because of the increase
of the collisional viscosity. The outer mass flux increases when the disk enters
rapidly the NSG regime ($r_{\rm p} \gtrsim 5\rm~m$), because of the increase of
the translational viscosity with $r_{\rm p}$ in the NSG regime. Here again the
intermediate state of the disk with 2.5 m-radius particles can be pointed out.

\subsection{Influence of the initial surface density profile}
In this section we study the influence of different initial surface-density
profiles. We modify one parameter at a time, and analyse its influence on the
viscous spreading over 5 billion years, and on the evolution of the disk's mass.

\subsubsection{Influence of initial mean radius}
First we change the initial position of the disk by shifting the surface density
maximum. While our standard model is defined with a mean radius of 110,000~km,
we study two other cases: 90,000~km and 130,000~km, which more or less covers
the total area of today's rings.

\begin{figure}[!h]
\begin{center}
\includegraphics[width=8.5cm]{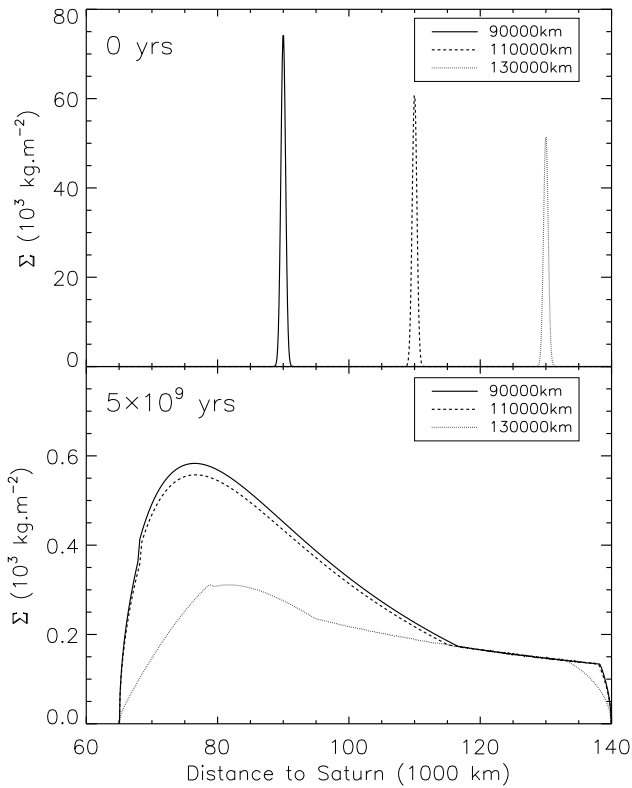}
\end{center}
\caption{Influence of initial position on surface density evolution.
\textbf{Top:} initial profiles. \textbf{Bottom:} at $t=5\rm~Gyrs$. When the
initial ring is farther from the planet it loses more mass, as it is then closer
to the Roche limit (see Fig. \ref{mass_dump_lr}).}
\label{pos_surf_dens}
\end{figure}

The surface density at $t=0$ and at 5 billion years is plotted in Fig.
\ref{pos_surf_dens}. The final disk masses are listed in Table
\ref{table_pos_mass}.

At 5 billion years, the disks starting at 90,000~km and 110,000~km have reached
a similar surface-density profile, and comparable disk mass. The mass fluxes
through the disk's boundaries have opposite behaviours: while the disk starting
at 90,000~km loses its mass preferentially through its inner boundary, the disk
starting at 110,000~km loses quite the same amount of mass but through the outer
boundary (Table \ref{table_pos_mass}).

The disk that starts at 130,000~km reaches the outer boundary very rapidly, in
about $10^5$ years, and loses a larger amount of mass than the other two. This
is due to the fact that when the disk is SG, the translational viscosity is
greatly increased at high radius, and also to the fact that the disk starts
closer to the Roche limit. As a result its final surface density is globally
lower than that of the other two disks. This disk never reaches the planet
within 5 billion years. 

\begin{table}[!h]
\caption{Influence of initial disk position on the evolution of the disk's
mass. Initial mass is $3.75\times10^{19}~\mbox{kg}$}
\begin{tabular}{l l l l}
\hline
Initial & Final & Mass through & Mass through\\
position & disk mass & inner edge & outer edge\\
(km) & $\left(10^{19}\rm~kg\right)$ & $\left(10^{19}\rm~kg\right)$ &
$\left(10^{19}\rm~kg\right)$\\
\hline
90,000 & 1,39 & 1,58 & 0,78\\
110,000 & 1.34 & 0.61 & 1.79\\
130,000 & 0.92 & $4\times10^{-4}$ & 2.83\\
\hline
\end{tabular}
\label{table_pos_mass}
\end{table}

Eventually, all disks still have a mass close to $10^{19}\rm~kg$, much
like in the standard model of section 3. The main effect of changing the disk's
initial position is to modify the material fluxes through the boundaries, and
increasing the initial radius increases the mass loss through the outer edge.

\subsubsection{Influence of initial width}
In this section we change the initial width of the disk by adjusting $\sigma$,
the standard deviation of the Gaussian. We compare the results of our standard
model to simulations with standard deviations $0.5\sigma$, $2\sigma$ and
$5\sigma$.

The modification of the initial width of the disk has no significant impact on
the evolution over 5 billion years. The most narrow disk, that starts with the
highest surface density, spreads more rapidly than the others and catches them
up in a few ten thousand years of evolution.

\subsubsection{Influence of initial mass repartition}
\begin{figure}[!h]
\begin{center}
\includegraphics[width=8.5cm]{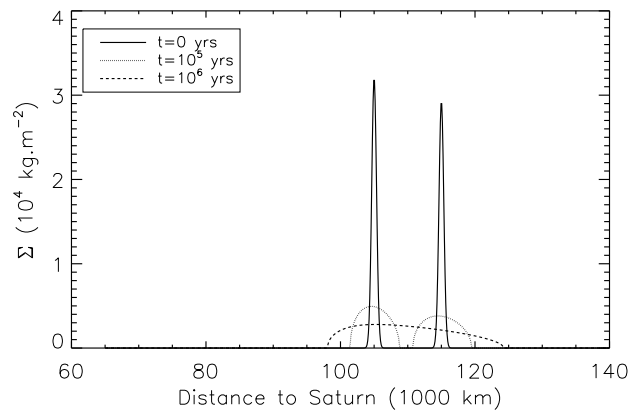}
\end{center}
\caption{Surface density evolution over the first million year for
an initial ring composed of two ringlets (plain line). Both ringlets contain the
same mass, but the outer one has lower surface densities because the ring's
surface ($\sim 2\pi RdR$) increases with distance. Both ringlets spread
similarly (dashed line). The two sub-rings merge in $\sim 10^6\rm~years$
(dotted line), after which the spreading continues as in the case with a single
initial ring.}
\label{dblgauss_surf_dens}
\end{figure}

Here we study a scenario where the initial material is distributed in two
ringlets instead of one. We assume a Gaussian profile for both of them and use
initial positions symmetrical with respect to R = 110,000~km. We choose to put
the exact same mass in both ringlets, so their initial maximal surface density
is different (because the surface of a bin increases with R). The two sub-rings
are distant of 10,000~km (peak to peak). The first million years of the
evolution of the disk is plotted in Fig. \ref{dblgauss_surf_dens}.

The two sub-rings merge in a few hundred thousand years, and then the evolution
goes on very similarly to the case with only one initial ring. The final mass of
the disk after 5 billion years of evolution is $1.34\times10^{19}\rm~kg$, very
close to what we obtained in section 3. The mass passing through the boundaries
is unaffected. The initial shape of the rings seems then to have very little
effect on their long-term evolution.

\section{Summary and discussion}
\subsection{Viscous age of the rings}

We have investigated the effect of a non-constant realistic viscosity model on
the global viscous spreading of dense planetary rings over 5 billion years. Our
main result is that using a physically realistic viscosity model introduces
important changes in the disk's spreading over time. We identified two
distinctively different spreading regimes, whether the disk is
self-gra\-vi\-ta\-ting or not. When the disk is self-gra\-vi\-ta\-ting, it
develops a surface density peak inward and an outward tail with lower surface
density and which is marginally gravitationally stable ($Q\sim2$), whereas a
totally non self-gra\-vi\-ta\-ting disk has a parabolic profile with a central
maximum. The disk's spreading is only significantly affected through
modification of the particle radius, because bigger particles increase the
Toomre's Q parameter.

Contrary to a continuous spreading of the disk with constant viscosity,
resulting in an emptying in $\sim500$ Myrs (depending on the constant viscosity
value), the spreading rate of the non-constant viscosity (NCV) disk
significantly decreases over time, allowing for a survival of an important part
of the disk over 5 billion years. We have shown in particular that with variable
viscosity in the self-gra\-vi\-ta\-ting regime, the disk's width increases like
$t^{1/4}$, much slower than in a constant viscosity case where the width
increases like $t^{1/2}$.

The final state of the disk seems somewhat independent of various initial
parameters such as initial width, position, and surface density profile.
Moreover, it appears that whatever the initial mass of the disk, the disk always
undergo a rapid initial emptying, which progressively slows down. We showed that
this evolution can be related to the disk's self-gra\-vi\-ta\-ting regime, with
emptying time-scales proportional to $1/M_0^2$ in the self-gra\-vi\-ta\-ting
regime, and $1/M_0$ in the non-self-gra\-vi\-ta\-ting regime. The disk's mass
drops continuously until it reaches the mass of a disk filling the area from the
planet's radius up to the Roche limit, and with a surface density so that every
part of the disk is marginally gravitationally stable ($Q \sim 2$), yielding
$\Sigma < 0.15\sqrt{M m_{\rm p}/r_{\rm p}} R^{-3/2}$. We computed that for the
specific case of Saturn this mass is $\sim 10^{19}~\rm kg$, in good agreement
with our simulation results (Fig. \ref{mass_init_influence}). Afterwards, the
disk evolves very slowly because of the very low viscosity in the NSG regime,
compared to the SG regime, for meter-sized particles.

In all our simulations starting with a disk mass about one Mimas' mass or more,
we could not achieve a total emptying of the rings in 5 billion years. Most of
the mass is lost through the Roche limit, except when the initial disk is
located below $\sim 100,000\rm~km$, where most of the mass is lost via infall
onto the planet. The survival of Saturn's rings against viscous spreading over
the age of the Solar System seems then possible, and during its evolution a lot
of material would have been made available for the accretion of satellites
\citep[][\textit{In press}]{char10}. 

The rings of Jupiter, Uranus and Neptune are today much less dense and massive
than Saturn's rings. Moreover, they are composed mainly of dust, while Saturn's
rings have typical particle sizes of $\sim 1\rm~cm$. For instance, the normal
optical depth of the Jovian ring system is $\sim 10^{-5}$, while the optical
depth of Saturn's A-ring is $\sim 0.1$ \citep{ocke99, ferr09}. In our
simulations, we showed that under viscous spreading an initially massive ring
system remains massive over time because of the very slowly evolving asymptotic
regime when the disk becomes marginally self-gra\-vi\-ta\-ting. Under these
considerations, it seems unlikely that the rings of the other giant planets
could have been dense enough to be self gra\-vi\-ta\-ting $\left(Q < 2\right)$
in their past.

\subsection{Comparison with observation}

\begin{figure}[!h]
\begin{center}
\includegraphics[width=8.5cm]{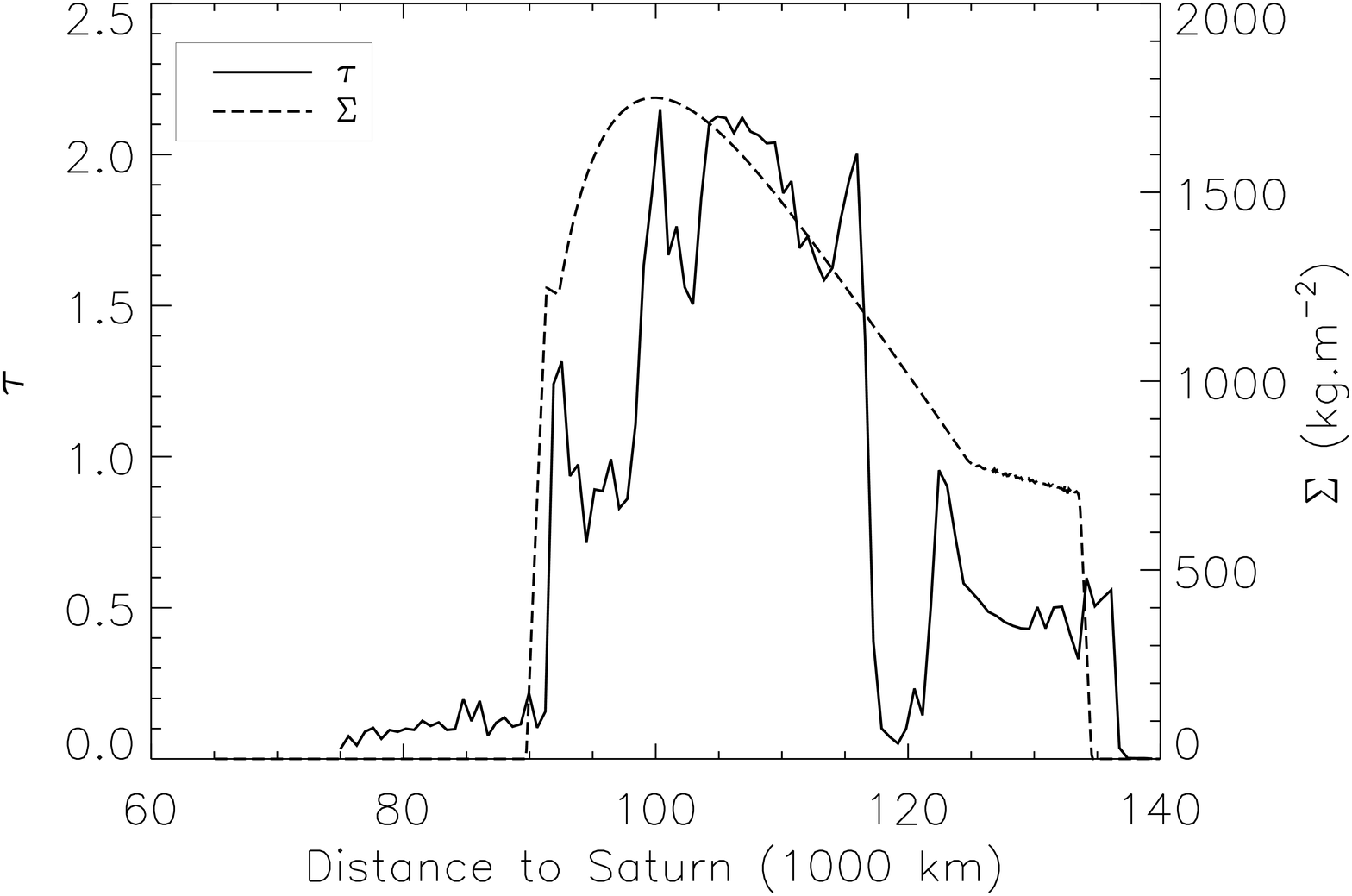}
\end{center}
\caption{Voyager PPS profile (plain line) and disk surface density at
t=$1.5\times10^8\rm~years$ for $r_{\rm p}=5\rm~m$ (dashed line). Both shapes are
similar, with a peak inward and lower densities outward.}
\label{pps}
\end{figure}

It is interesting to compare the current rings of Saturn with the
surface-density profiles we obtain in our simulations. We have plotted in Fig.
\ref{pps} the ring optical depth $\tau$ from the Voyager photo-polarimeter
spectrum (PPS), as it may be proportional to the surface mass density in low
$\tau$ regions. We have also plotted the disk surface density at
$t=1.5\times10^8\rm~years$ for the disk with 5 m-radius particles. Apart from
the Cassini Division, which is thought to have been created by Mimas' 2:1
resonance \citep{gold78} and the C ring, we note several similarities. 

\begin{figure*}[!ht]
\begin{center}
\includegraphics[width=16cm]{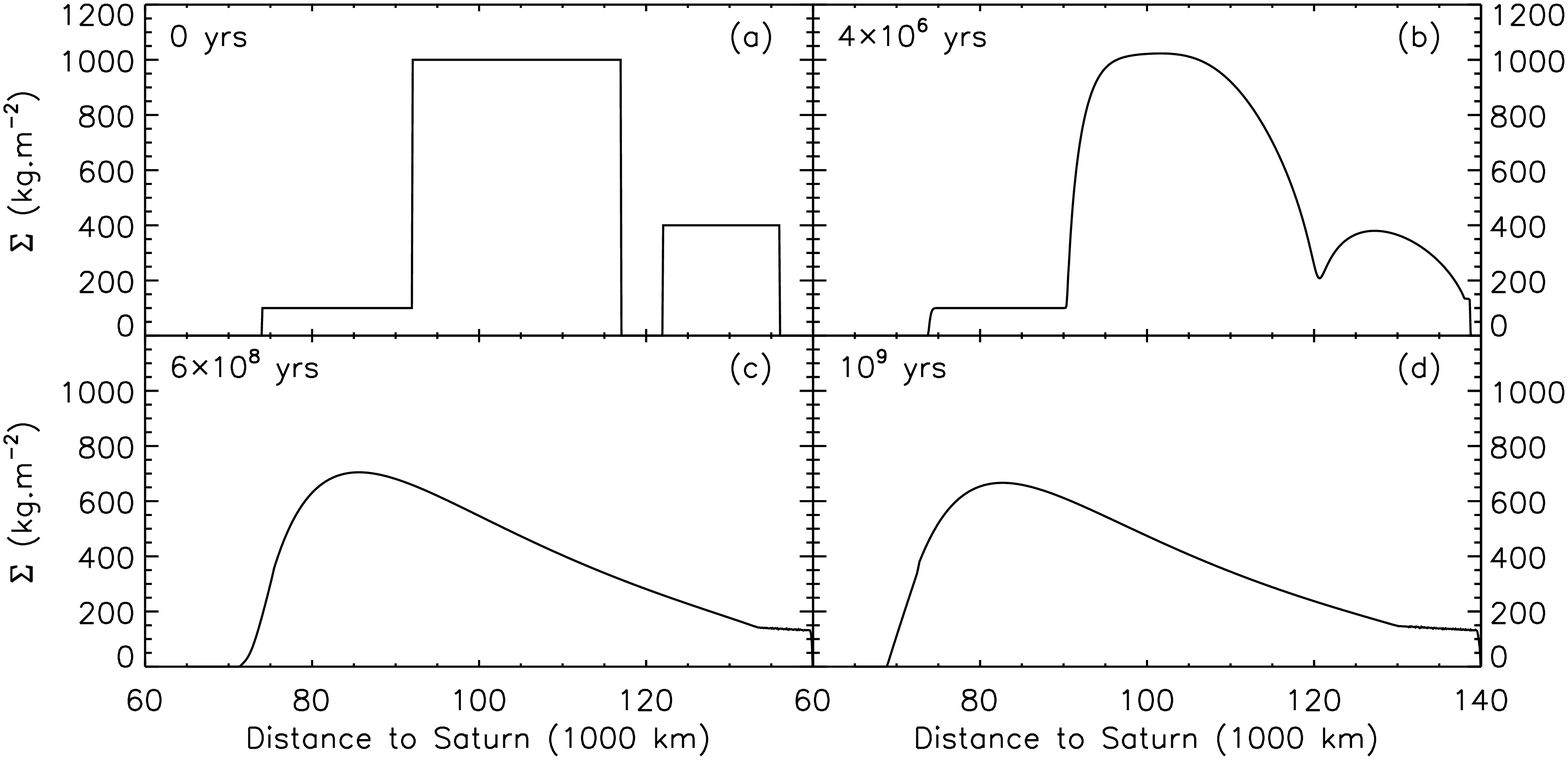}
\end{center}
\caption{Current rings surface density evolution. \textbf{(a)} 
Initial profile. \textbf{(b)} At 4 Myrs of evolution. \textbf{(c)} At 600 Myrs
of evolution. \textbf{(d)} At 1 Gyrs of evolution. The B ring rapidly fills the
Cassini Division, and overwhelms the C ring in $\sim 1\rm~Gyrs$. As in the
previous simulations, the disk evolves with a peak inward and a marginally
self-gra\-vi\-ta\-ting outer plateau (see Fig. \ref{sg_trans_surf_dens}).}
\label{cur_rings}
\end{figure*}

Firstly, the global shape is in good agreement, with a maximum inward and lower
surface densities toward the outer limits of the rings, with the peak
corresponding to the B ring and the outer plateau to the A ring. This plateau
is marginally gravitationally stable ($Q \sim 2$), which is coherent with the
several observations of wakes in the A ring \citep{ferr09}. The position of the
maximum is much more inward later in our simulations at $\sim10^9\rm~years$.
This favours the hypothesis that Saturn's rings are young, but this could also
be explained by the imperfections of the viscosity model we use (the mass has
been transferred inward too rapidly), or also to other physical effects we have
not considered in our study, or to unknown initial conditions. A change in the
particle radius would also modify significantly the optical depth profile.

We can note also that in our simulations we never recover anything similar to
the C ring. This could be due to an important modification of the particle size
in this region, but this is unlikely as particle size distributions show a
greater power index for the C-ring than for the B-ring, so that there should be
less large particles in the C-ring than in the B-ring \citep[Section
17.2.1.1]{char09b}. Moreover this modification of the ring particle size would
not explain the very steep transition from the B ring to the C ring. This
transition may however suggest that the C ring is a product of another physical
process. For instance, meteoritic bombardment is though to be capable of
projecting large amounts of material over significant distances \citep{cuzz98}.
The C-ring could then be ejecta from the bombardment of the B-ring. This would
in turn agree with the observed similarities in the spectral signatures of the B
and C-rings as observed by Cassini VIMS \citep{nich08}

Although today's rings mostly resemble a 100 million-years old disk when
compared to our simulations, their formation so recently in the history of the
Solar System is yet to be explained. In this work we have considered only the
viscous torque, so that we can only say that today's rings \textit{viscously}
look like a 100 million-years old disk. Other important physical effects have to
be included in order to fully constrain the evolution time-scales of the rings.
For instance, a satellite located between the planet and the initial rings could
importantly slow down the inward spreading of the rings via resonant
interactions, which apply a repulsive torque to the disk \citep{meye87}.
Conversely, outer satellites could also substantially reduce the rings outward
spreading. These interactions would consequently reduce the mass lost by the
disk through its outer boundary. Considering that in our simulations, most of
the mass is lost through this boundary, satellite interactions appear as a
process that could significantly increase the survival of the rings. However
these interactions should increase the mass lost by the disk from its inner
boundary, because of the repulsive torque applied by the satellite on the disk.
Inclusion of satellite interactions would then be important to quantify the
impact on the mass lost by the disk.

\subsection{Viscous spreading of current rings}

While the aim of our work was not to reproduce today's rings but to understand
the effects of a realistic viscosity model over long time-scales, we applied our
model to an initial set of parameters that roughly reproduces today's rings of
Saturn: a first slab from 74,000~km to 92,000~km with $\Sigma =
100~\rm kg.m^{-2}$ (the C ring), a second slab from 92,000~km to 117,000~km 
with
$\Sigma = 1000~\rm kg.m^{-2}$ (the B ring), a gap from 117,000~km to 122,000~km
(the Cassini Division) and a third slab from 122,000~km to 136,000~km with $
\Sigma = 400~\rm kg.m^{-2}$ (the A ring). We choose a particle radius of 1m.
Results
are plotted in Fig. \ref{cur_rings}.

In this configuration, the total rings' mass is $2.2\times10^{19}\rm~kg$ and
only the A and B rings are self-gra\-vi\-ta\-ting, the C ring being not dense
enough. The average initial viscosities are $10^{-4}\rm~m^2.s^{-1}$ for the C
ring, and $10^{-2}\rm~m^2.s^{-1}$ for the B and A rings, which is in good
agreement with observational results \citep{tisc07}.

The Cassini Division is filled in about 4 million years, because of a rapid
outward spreading of the B ring (Fig. \ref{cur_rings}b). This emphasizes the
role of the 2:1 Mimas resonance in the existence of the Cassini Division
\citep{schw81, espo87}. The inner spreading of the B ring is much slower. The C
ring, because of its very low viscosity, evolves very little. It is completely
``eaten'' by the B ring in approximately 600 million years (Fig.
\ref{cur_rings}c).

It is interesting to note that the shape of the surface density at $10^9$ years
is very similar to what we obtain with our standard model (Fig.
\ref{rp_influence_surf_dens}c-d, light-blue curve, and \ref{cur_rings}d), even
though the initial surface density profiles where very different in the two
simulations. The final state of the rings seems then very independent of the
initial conditions.

After 1 billion years of evolution, the mass of the disk is still
$1.6\times10^{19}\rm~kg$. It seems then, as we have seen in section 3, that
Saturn's rings could very well be viscously old, but they are also not likely to
disappear in the next billion years.

\subsection{Perspectives}
An important improvement to this work would be to include a particle size
distribution in the viscosity model. Indeed, we have seen in section 4 that only
the ring-particle radius significantly affects the spreading of the rings as it
impacts the positions of the transitions between the self-gra\-vi\-ta\-ting and
non-self-gra\-vi\-ta\-ting regimes, for a given surface density. This means that
if particle segregation occurs in the rings, this could lead to modification of
the viscous behaviour of some regions of the rings.

While viscous spreading is a key physical phe\-no\-me\-non for the evolution
of Saturn's rings, it will be affected by other effects such as resonant torques
with satellite. Many structures of the rings are known to be the result of
resonant interactions with the satellites of Saturn \citep{espo87}. The
repulsive torque exerted by outer satellites could also significantly alter the
spreading of the disk \citep{salm09}

These structures, for instance density waves, locally modify the density and
thus the viscosity. How material is transported through these waves could gives
us clues on the suggested ``recycling'' of the rings, which could be responsible
for the apparent low meteoritic pollution of the rings \citep{cuzz98}. It would
also be important to constrain the long term influence of these resonant
interactions, and see if they increase the possible lifetime of Saturn's rings
or if they contribute to their destruction.

\section*{Acknowledgements}
We would like to thank Dr. Anais Rassat for her careful reading of this paper
and her very useful comments and suggestions. We would like also to thank
Pierre-Yves Longaretti, and an anonymous referee, for their very useful review
that contributed to increase the quality of this paper. This work was supported
by the Commissariat \`a l'\'Energie Atomique and Universit\'e Paris Diderot.

\bibliography{biblio}

\begin{thebibliography}{47}
\expandafter\ifx\csname natexlab\endcsname\relax\def\natexlab#1{#1}\fi
\expandafter\ifx\csname url\endcsname\relax
  \def\url#1{\texttt{#1}}\fi
\expandafter\ifx\csname urlprefix\endcsname\relax\def\urlprefix{URL }\fi
\providecommand{\eprint}[2][]{\url{#2}}
\providecommand{\bibinfo}[2]{#2}
\ifx\xfnm\relax \def\xfnm[#1]{\unskip,\space#1}\fi
\bibitem[{{Albers} and {Spahn}(2006)}]{albe06}
\bibinfo{author}{{Albers}, N.}, \bibinfo{author}{{Spahn}, F.},
  \bibinfo{year}{2006}.
\newblock \bibinfo{title}{{The influence of particle adhesion on the stability
  of agglomerates in Saturn's rings}}.
\newblock \bibinfo{journal}{Icarus} \bibinfo{volume}{181},
  \bibinfo{pages}{292--301}.
\bibitem[{{Alibert} et~al.(2005){Alibert}, {Mousis} and {Benz}}]{alib05}
\bibinfo{author}{{Alibert}, Y.}, \bibinfo{author}{{Mousis}, O.},
  \bibinfo{author}{{Benz}, W.}, \bibinfo{year}{2005}.
\newblock \bibinfo{title}{{Modeling the Jovian subnebula. I. Thermodynamic
  conditions and migration of proto-satellites}}.
\newblock \bibinfo{journal}{A\&A} \bibinfo{volume}{439},
  \bibinfo{pages}{1205--1213}.
\newblock \eprint{arXiv:astro-ph/0505367}.
\bibitem[{{Araki} and {Tremaine}(1986)}]{arak86}
\bibinfo{author}{{Araki}, S.}, \bibinfo{author}{{Tremaine}, S.},
  \bibinfo{year}{1986}.
\newblock \bibinfo{title}{{The dynamics of dense particle disks}}.
\newblock \bibinfo{journal}{Icarus} \bibinfo{volume}{65},
  \bibinfo{pages}{83--109}.
\bibitem[{{Baruteau} and {Masset}(2008)}]{baru08}
\bibinfo{author}{{Baruteau}, C.}, \bibinfo{author}{{Masset}, F.},
  \bibinfo{year}{2008}.
\newblock \bibinfo{title}{{Type I Planetary migration in a self-gravitating
  disk}}.
\newblock \bibinfo{journal}{ApJ} \bibinfo{volume}{678},
  \bibinfo{pages}{483--497}.
\newblock \eprint{0801.4413}.
\bibitem[{{Bath} and {Pringle}(1981)}]{bath81}
\bibinfo{author}{{Bath}, G.T.}, \bibinfo{author}{{Pringle}, J.E.},
  \bibinfo{year}{1981}.
\newblock \bibinfo{title}{{The evolution of viscous discs. I - Mass transfer
  variations}}.
\newblock \bibinfo{journal}{MNRAS} \bibinfo{volume}{194},
  \bibinfo{pages}{967--986}.
\bibitem[{{Charnoz} et~al.(2009a){Charnoz}, {Dones}, {Esposito}, {Estrada} and
  {Hedman}}]{char09b}
\bibinfo{author}{{Charnoz}, S.}, \bibinfo{author}{{Dones}, L.},
  \bibinfo{author}{{Esposito}, L.W.}, \bibinfo{author}{{Estrada}, P.R.},
  \bibinfo{author}{{Hedman}, M.M.}, \bibinfo{year}{2009a}.
\newblock \bibinfo{title}{{Origin and evolution of Saturn's ring system}}.
\newblock pp. \bibinfo{pages}{537--573}.
\bibitem[{{Charnoz} et~al.(2009b){Charnoz}, {Morbidelli}, {Dones} and
  {Salmon}}]{char09a}
\bibinfo{author}{{Charnoz}, S.}, \bibinfo{author}{{Morbidelli}, A.},
  \bibinfo{author}{{Dones}, L.}, \bibinfo{author}{{Salmon}, J.},
  \bibinfo{year}{2009b}.
\newblock \bibinfo{title}{{Did Saturn's rings form during the Late Heavy
  Bombardment?}}
\newblock \bibinfo{journal}{Icarus} \bibinfo{volume}{199},
  \bibinfo{pages}{413--428}.
\newblock \eprint{0809.5073}.
\bibitem[{{Charnoz} et~al.(){Charnoz}, {Salmon} and {Crida}}]{char10}
\bibinfo{author}{{Charnoz}, S.}, \bibinfo{author}{{Salmon}, J.},
  \bibinfo{author}{{Crida}, A.}, .
\newblock \bibinfo{title}{{The recent formation of Saturn's moonlets from
  viscous spreading of the main rings}}.
\newblock \bibinfo{note}{Nature, \textit{In press}}.
\bibitem[{{Cuzzi} and {Estrada}(1998)}]{cuzz98}
\bibinfo{author}{{Cuzzi}, J.N.}, \bibinfo{author}{{Estrada}, P.R.},
  \bibinfo{year}{1998}.
\newblock \bibinfo{title}{{Compositional evolution of Saturn's rings due to
  meteoroid bombardment}}.
\newblock \bibinfo{journal}{Icarus} \bibinfo{volume}{132},
  \bibinfo{pages}{1--35}.
\bibitem[{{Cuzzi} and {Pollack}(1978)}]{cuzz78}
\bibinfo{author}{{Cuzzi}, J.N.}, \bibinfo{author}{{Pollack}, J.B.},
  \bibinfo{year}{1978}.
\newblock \bibinfo{title}{{Saturn's rings: particle composition and size
  distribution as constrained by microwave observations. I - Radar
  observations}}.
\newblock \bibinfo{journal}{Icarus} \bibinfo{volume}{33},
  \bibinfo{pages}{233--262}.
\bibitem[{{Cuzzi} et~al.(1980){Cuzzi}, {Pollack} and {Summers}}]{cuzz80}
\bibinfo{author}{{Cuzzi}, J.N.}, \bibinfo{author}{{Pollack}, J.B.},
  \bibinfo{author}{{Summers}, A.L.}, \bibinfo{year}{1980}.
\newblock \bibinfo{title}{{Saturn's rings - particle composition and size
  distribution as constrained by observations at microwave wavelengths. II -
  Radio interferometric observations}}.
\newblock \bibinfo{journal}{Icarus} \bibinfo{volume}{44},
  \bibinfo{pages}{683--705}.
\bibitem[{{Daisaka} and {Ida}(1999)}]{dais99}
\bibinfo{author}{{Daisaka}, H.}, \bibinfo{author}{{Ida}, S.},
  \bibinfo{year}{1999}.
\newblock \bibinfo{title}{{Spatial structure and coherent motion in dense
  planetary rings induced by self-gravitational instability}}.
\newblock \bibinfo{journal}{Earth, Planets, and Space} \bibinfo{volume}{51},
  \bibinfo{pages}{1195--1213}.
\newblock \eprint{arXiv:astro-ph/9908057}.
\bibitem[{{Daisaka} et~al.(2001){Daisaka}, {Tanaka} and {Ida}}]{dais01}
\bibinfo{author}{{Daisaka}, H.}, \bibinfo{author}{{Tanaka}, H.},
  \bibinfo{author}{{Ida}, S.}, \bibinfo{year}{2001}.
\newblock \bibinfo{title}{{Viscosity in a dense planetary ring with
  self-gravitating particles}}.
\newblock \bibinfo{journal}{Icarus} \bibinfo{volume}{154},
  \bibinfo{pages}{296--312}.
\bibitem[{{Dones}(1991)}]{done91}
\bibinfo{author}{{Dones}, L.}, \bibinfo{year}{1991}.
\newblock \bibinfo{title}{{A recent cometary origin for Saturn's rings?}}
\newblock \bibinfo{journal}{Icarus} \bibinfo{volume}{92},
  \bibinfo{pages}{194--203}.
\bibitem[{{Esposito}(1986)}]{espo86}
\bibinfo{author}{{Esposito}, L.W.}, \bibinfo{year}{1986}.
\newblock \bibinfo{title}{{Structure and evolution of Saturn's rings}}.
\newblock \bibinfo{journal}{Icarus} \bibinfo{volume}{67},
  \bibinfo{pages}{345--357}.
\bibitem[{{Esposito} et~al.(1987){Esposito}, {Harris} and {Simmons}}]{espo87}
\bibinfo{author}{{Esposito}, L.W.}, \bibinfo{author}{{Harris}, C.C.},
  \bibinfo{author}{{Simmons}, K.E.}, \bibinfo{year}{1987}.
\newblock \bibinfo{title}{{Features in Saturn's rings}}.
\newblock \bibinfo{journal}{ApJS} \bibinfo{volume}{63},
  \bibinfo{pages}{749--770}.
\bibitem[{{Esposito} et~al.(1983){Esposito}, {Ocallaghan} and {West}}]{espo83}
\bibinfo{author}{{Esposito}, L.W.}, \bibinfo{author}{{Ocallaghan}, M.},
  \bibinfo{author}{{West}, R.A.}, \bibinfo{year}{1983}.
\newblock \bibinfo{title}{{The structure of Saturn's rings - Implications from
  the Voyager stellar occultation}}.
\newblock \bibinfo{journal}{Icarus} \bibinfo{volume}{56},
  \bibinfo{pages}{439--452}.
\bibitem[{{Ferrari} et~al.(2009){Ferrari}, {Brooks}, {Edgington}, {Leyrat},
  {Pilorz} and {Spilker}}]{ferr09}
\bibinfo{author}{{Ferrari}, C.}, \bibinfo{author}{{Brooks}, S.},
  \bibinfo{author}{{Edgington}, S.}, \bibinfo{author}{{Leyrat}, C.},
  \bibinfo{author}{{Pilorz}, S.}, \bibinfo{author}{{Spilker}, L.},
  \bibinfo{year}{2009}.
\newblock \bibinfo{title}{{Structure of self-gravity wakes in Saturn's A ring
  as measured by Cassini CIRS}}.
\newblock \bibinfo{journal}{Icarus} \bibinfo{volume}{199},
  \bibinfo{pages}{145--153}.
\bibitem[{{Goldreich} and {Tremaine}(1978)}]{gold78}
\bibinfo{author}{{Goldreich}, P.}, \bibinfo{author}{{Tremaine}, S.},
  \bibinfo{year}{1978}.
\newblock \bibinfo{title}{{The formation of the Cassini division in Saturn's
  rings}}.
\newblock \bibinfo{journal}{Icarus} \bibinfo{volume}{34},
  \bibinfo{pages}{240--253}.
\bibitem[{{Goldreich} and {Tremaine}(1979)}]{gold79}
\bibinfo{author}{{Goldreich}, P.}, \bibinfo{author}{{Tremaine}, S.},
  \bibinfo{year}{1979}.
\newblock \bibinfo{title}{{The excitation of density waves at the Lindblad and
  corotation resonances by an external potential}}.
\newblock \bibinfo{journal}{ApJ} \bibinfo{volume}{233},
  \bibinfo{pages}{857--871}.
\bibitem[{{Goldreich} and {Tremaine}(1982)}]{gold82}
\bibinfo{author}{{Goldreich}, P.}, \bibinfo{author}{{Tremaine}, S.},
  \bibinfo{year}{1982}.
\newblock \bibinfo{title}{{The dynamics of planetary rings}}.
\newblock \bibinfo{journal}{ARA\&A} \bibinfo{volume}{20},
  \bibinfo{pages}{249--283}.
\bibitem[{{Harris}(1984)}]{harr84}
\bibinfo{author}{{Harris}, A.W.}, \bibinfo{year}{1984}.
\newblock \bibinfo{title}{{The origin and evolution of planetary rings}}, in:
  \bibinfo{editor}{{R.~Greenberg \& A.~Brahic}} (Ed.), \bibinfo{booktitle}{IAU
  Colloq. 75: Planetary Rings}, pp. \bibinfo{pages}{641--659}.
\bibitem[{{Larson}(1984)}]{lars84}
\bibinfo{author}{{Larson}, R.B.}, \bibinfo{year}{1984}.
\newblock \bibinfo{title}{{Gravitational torques and star formation}}.
\newblock \bibinfo{journal}{MNRAS} \bibinfo{volume}{206},
  \bibinfo{pages}{197--207}.
\bibitem[{{Latter} and {Ogilvie}(2009)}]{latt09}
\bibinfo{author}{{Latter}, H.N.}, \bibinfo{author}{{Ogilvie}, G.I.},
  \bibinfo{year}{2009}.
\newblock \bibinfo{title}{{The viscous overstability, nonlinear wavetrains, and
  finescale structure in dense planetary rings}}.
\newblock \bibinfo{journal}{Icarus} \bibinfo{volume}{202},
  \bibinfo{pages}{565--583}.
\newblock \eprint{0904.0143}.
\bibitem[{{Lin} and {Pringle}(1987)}]{lin87}
\bibinfo{author}{{Lin}, D.N.C.}, \bibinfo{author}{{Pringle}, J.E.},
  \bibinfo{year}{1987}.
\newblock \bibinfo{title}{{A viscosity prescription for a self-gravitating
  accretion disc}}.
\newblock \bibinfo{journal}{MNRAS} \bibinfo{volume}{225},
  \bibinfo{pages}{607--613}.
\bibitem[{{Lynden-Bell} and {Pringle}(1974)}]{lynd74}
\bibinfo{author}{{Lynden-Bell}, D.}, \bibinfo{author}{{Pringle}, J.E.},
  \bibinfo{year}{1974}.
\newblock \bibinfo{title}{{The evolution of viscous discs and the origin of the
  nebular variables.}}
\newblock \bibinfo{journal}{MNRAS} \bibinfo{volume}{168},
  \bibinfo{pages}{603--637}.
\bibitem[{{Marouf} et~al.(1983){Marouf}, {Tyler}, {Zebker}, {Simpson} and
  {Eshleman}}]{maro83}
\bibinfo{author}{{Marouf}, E.A.}, \bibinfo{author}{{Tyler}, G.L.},
  \bibinfo{author}{{Zebker}, H.A.}, \bibinfo{author}{{Simpson}, R.A.},
  \bibinfo{author}{{Eshleman}, V.R.}, \bibinfo{year}{1983}.
\newblock \bibinfo{title}{{Particle size distributions in Saturn's rings from
  Voyager 1 radio occultation}}.
\newblock \bibinfo{journal}{Icarus} \bibinfo{volume}{54},
  \bibinfo{pages}{189--211}.
\bibitem[{{Masset} and {Casoli}(2009)}]{mass09}
\bibinfo{author}{{Masset}, F.S.}, \bibinfo{author}{{Casoli}, J.},
  \bibinfo{year}{2009}.
\newblock \bibinfo{title}{{On the horseshoe drag of a low-mass planet. II.
  Migration in adiabatic disks}}.
\newblock \bibinfo{journal}{ApJ} \bibinfo{volume}{703},
  \bibinfo{pages}{857--876}.
\newblock \eprint{0907.4676}.
\bibitem[{{Meyer-Vernet} and {Sicardy}(1987)}]{meye87}
\bibinfo{author}{{Meyer-Vernet}, N.}, \bibinfo{author}{{Sicardy}, B.},
  \bibinfo{year}{1987}.
\newblock \bibinfo{title}{{On the physics of resonant disk-satellite
  interaction}}.
\newblock \bibinfo{journal}{Icarus} \bibinfo{volume}{69},
  \bibinfo{pages}{157--175}.
\bibitem[{{Nicholson} et~al.(2008){Nicholson}, {Hedman}, {Clark}, {Showalter},
  {Cruikshank}, {Cuzzi}, {Filacchione}, {Capaccioni}, {Cerroni}, {Hansen},
  {Sicardy}, {Drossart}, {Brown}, {Buratti}, {Baines} and {Coradini}}]{nich08}
\bibinfo{author}{{Nicholson}, P.D.}, \bibinfo{author}{{Hedman}, M.M.},
  \bibinfo{author}{{Clark}, R.N.}, \bibinfo{author}{{Showalter}, M.R.},
  \bibinfo{author}{{Cruikshank}, D.P.}, \bibinfo{author}{{Cuzzi}, J.N.},
  \bibinfo{author}{{Filacchione}, G.}, \bibinfo{author}{{Capaccioni}, F.},
  \bibinfo{author}{{Cerroni}, P.}, \bibinfo{author}{{Hansen}, G.B.},
  \bibinfo{author}{{Sicardy}, B.}, \bibinfo{author}{{Drossart}, P.},
  \bibinfo{author}{{Brown}, R.H.}, \bibinfo{author}{{Buratti}, B.J.},
  \bibinfo{author}{{Baines}, K.H.}, \bibinfo{author}{{Coradini}, A.},
  \bibinfo{year}{2008}.
\newblock \bibinfo{title}{{A close look at Saturn's rings with Cassini VIMS}}.
\newblock \bibinfo{journal}{Icarus} \bibinfo{volume}{193},
  \bibinfo{pages}{182--212}.
\bibitem[{{Ockert-Bell} et~al.(1999){Ockert-Bell}, {Burns}, {Daubar}, {Thomas},
  {Veverka}, {Belton} and {Klaasen}}]{ocke99}
\bibinfo{author}{{Ockert-Bell}, M.E.}, \bibinfo{author}{{Burns}, J.A.},
  \bibinfo{author}{{Daubar}, I.J.}, \bibinfo{author}{{Thomas}, P.C.},
  \bibinfo{author}{{Veverka}, J.}, \bibinfo{author}{{Belton}, M.J.S.},
  \bibinfo{author}{{Klaasen}, K.P.}, \bibinfo{year}{1999}.
\newblock \bibinfo{title}{{The structure of Jupiter's ring system as revealed
  by the Galileo Imaging Experiment}}.
\newblock \bibinfo{journal}{Icarus} \bibinfo{volume}{138},
  \bibinfo{pages}{188--213}.
\bibitem[{{Ohtsuki}(1999)}]{ohts99}
\bibinfo{author}{{Ohtsuki}, K.}, \bibinfo{year}{1999}.
\newblock \bibinfo{title}{{Evolution of particle velocity dispersion in a
  circumplanetary disk due to inelastic collisions and gravitational
  interactions}}.
\newblock \bibinfo{journal}{Icarus} \bibinfo{volume}{137},
  \bibinfo{pages}{152--177}.
\bibitem[{{Ohtsuki} and {Emori}(2000)}]{ohts00}
\bibinfo{author}{{Ohtsuki}, K.}, \bibinfo{author}{{Emori}, H.},
  \bibinfo{year}{2000}.
\newblock \bibinfo{title}{{Local N-body simulations for the distribution and
  evolution of particle velocities in planetary rings}}.
\newblock \bibinfo{journal}{AJ} \bibinfo{volume}{119},
  \bibinfo{pages}{403--416}.
\bibitem[{{Pollack} et~al.(1973){Pollack}, {Summers} and {Baldwin}}]{poll73}
\bibinfo{author}{{Pollack}, J.B.}, \bibinfo{author}{{Summers}, A.},
  \bibinfo{author}{{Baldwin}, B.}, \bibinfo{year}{1973}.
\newblock \bibinfo{title}{{Estimates of the sizes of the particles in the rings
  of Saturn and their cosmogonic implications}}.
\newblock \bibinfo{journal}{Icarus} \bibinfo{volume}{20},
  \bibinfo{pages}{263--278}.
\bibitem[{{Porco} et~al.(2008){Porco}, {Weiss}, {Richardson}, {Dones}, {Quinn}
  and {Throop}}]{porc08}
\bibinfo{author}{{Porco}, C.C.}, \bibinfo{author}{{Weiss}, J.W.},
  \bibinfo{author}{{Richardson}, D.C.}, \bibinfo{author}{{Dones}, L.},
  \bibinfo{author}{{Quinn}, T.}, \bibinfo{author}{{Throop}, H.},
  \bibinfo{year}{2008}.
\newblock \bibinfo{title}{{Simulations of the dynamical and light-scattering
  behavior of saturn's rings and the derivation of ring particle and disk
  properties}}.
\newblock \bibinfo{journal}{AJ} \bibinfo{volume}{136},
  \bibinfo{pages}{2172--2200}.
\bibitem[{{Pringle}(1981)}]{prin81}
\bibinfo{author}{{Pringle}, J.E.}, \bibinfo{year}{1981}.
\newblock \bibinfo{title}{{Accretion discs in astrophysics}}.
\newblock \bibinfo{journal}{ARA\&A} \bibinfo{volume}{19},
  \bibinfo{pages}{137--162}.
\bibitem[{{Richardson}(1994)}]{rich94}
\bibinfo{author}{{Richardson}, D.C.}, \bibinfo{year}{1994}.
\newblock \bibinfo{title}{{Tree code simulations of planetary rings}}.
\newblock \bibinfo{journal}{MNRAS} \bibinfo{volume}{269},
  \bibinfo{pages}{493--+}.
\bibitem[{{Salmon} et~al.(2009){Salmon}, {Charnoz}, {Crida} and
  {Brahic}}]{salm09}
\bibinfo{author}{{Salmon}, J.}, \bibinfo{author}{{Charnoz}, S.},
  \bibinfo{author}{{Crida}, A.}, \bibinfo{author}{{Brahic}, A.},
  \bibinfo{year}{2009}.
\newblock \bibinfo{title}{{Simulations of Saturn's rings evolution over 5
  billion years with variable viscosity {\&} satellite interactions}}, in:
  \bibinfo{booktitle}{AAS/Division for Planetary Sciences Meeting Abstracts},
  pp. \bibinfo{pages}{\#25.08--+}.
\bibitem[{{Salo}(1992)}]{salo92}
\bibinfo{author}{{Salo}, H.}, \bibinfo{year}{1992}.
\newblock \bibinfo{title}{{Gravitational wakes in Saturn's rings}}.
\newblock \bibinfo{journal}{Nature} \bibinfo{volume}{359},
  \bibinfo{pages}{619--621}.
\bibitem[{{Salo}(1995)}]{salo95}
\bibinfo{author}{{Salo}, H.}, \bibinfo{year}{1995}.
\newblock \bibinfo{title}{{Simulations of dense planetary rings. III.
  Self-gravitating identical particles.}}
\newblock \bibinfo{journal}{Icarus} \bibinfo{volume}{117},
  \bibinfo{pages}{287--312}.
\bibitem[{{Schwartz}(1981)}]{schw81}
\bibinfo{author}{{Schwartz}, M.P.}, \bibinfo{year}{1981}.
\newblock \bibinfo{title}{{Clearing the Cassini division}}.
\newblock \bibinfo{journal}{Icarus} \bibinfo{volume}{48},
  \bibinfo{pages}{339--342}.
\bibitem[{{Shu} and {Stewart}(1985)}]{shu85}
\bibinfo{author}{{Shu}, F.H.}, \bibinfo{author}{{Stewart}, G.R.},
  \bibinfo{year}{1985}.
\newblock \bibinfo{title}{{The collisional dynamics of particulate disks}}.
\newblock \bibinfo{journal}{Icarus} \bibinfo{volume}{62},
  \bibinfo{pages}{360--383}.
\bibitem[{{Takeda} and {Ida}(2001)}]{take01}
\bibinfo{author}{{Takeda}, T.}, \bibinfo{author}{{Ida}, S.},
  \bibinfo{year}{2001}.
\newblock \bibinfo{title}{{Angular momentum transfer in a protolunar disk}}.
\newblock \bibinfo{journal}{ApJ} \bibinfo{volume}{560},
  \bibinfo{pages}{514--533}.
\newblock \eprint{arXiv:astro-ph/0108133}.
\bibitem[{{Takeuchi} et~al.(1996){Takeuchi}, {Miyama} and {Lin}}]{take96}
\bibinfo{author}{{Takeuchi}, T.}, \bibinfo{author}{{Miyama}, S.M.},
  \bibinfo{author}{{Lin}, D.N.C.}, \bibinfo{year}{1996}.
\newblock \bibinfo{title}{{Gap formation in protoplanetary disks}}.
\newblock \bibinfo{journal}{ApJ} \bibinfo{volume}{460},
  \bibinfo{pages}{832--+}.
\bibitem[{{Tiscareno} et~al.(2007){Tiscareno}, {Burns}, {Nicholson}, {Hedman}
  and {Porco}}]{tisc07}
\bibinfo{author}{{Tiscareno}, M.S.}, \bibinfo{author}{{Burns}, J.A.},
  \bibinfo{author}{{Nicholson}, P.D.}, \bibinfo{author}{{Hedman}, M.M.},
  \bibinfo{author}{{Porco}, C.C.}, \bibinfo{year}{2007}.
\newblock \bibinfo{title}{{Cassini imaging of Saturn's rings. II. A wavelet
  technique for analysis of density waves and other radial structure in the
  rings}}.
\newblock \bibinfo{journal}{Icarus} \bibinfo{volume}{189},
  \bibinfo{pages}{14--34}.
\newblock \eprint{arXiv:astro-ph/0610242}.
\bibitem[{{Toomre}(1964)}]{toom64}
\bibinfo{author}{{Toomre}, A.}, \bibinfo{year}{1964}.
\newblock \bibinfo{title}{{On the gravitational stability of a disk of stars}}.
\newblock \bibinfo{journal}{ApJ} \bibinfo{volume}{139},
  \bibinfo{pages}{1217--1238}.
\bibitem[{{Wisdom} and {Tremaine}(1988)}]{wisd88}
\bibinfo{author}{{Wisdom}, J.}, \bibinfo{author}{{Tremaine}, S.},
  \bibinfo{year}{1988}.
\newblock \bibinfo{title}{{Local simulations of planetary rings}}.
\newblock \bibinfo{journal}{AJ} \bibinfo{volume}{95},
  \bibinfo{pages}{925--940}.

\end{thebibliography}

\end{document}